\documentclass[11pt,aip,jcp,amsmath,amssymb]{revtex4-1}

\usepackage{graphicx,bm,times,hyperref}

\begin{document}

\title{Enhanced solvation force extrapolation for speeding up molecular dynamics simulations of complex biochemical liquids}

\author{Igor Omelyan}
\email{omelyan@icmp.lviv.ua}
\affiliation{Institute for Condensed Matter Physics,
National Academy of Sciences of Ukraine, 1 Svientsitskii
Street, Lviv 79011; Ukraine}

\author{Andriy Kovalenko}
\affiliation{Department of Mechanical Engineering, University of
Alberta, 9211-116 Street NW, Edmonton, AB T6G 1H9, Canada}

\affiliation{Nanotechnology Research Centre, 11421 Saskatchewan Drive,
Edmonton, AB T6G 2M9; Canada}

\date{\today}

\begin{abstract}

We propose an enhanced approach to the extrapolation of mean potential forces acting on atoms of solute macromolecules due to their interactions with solvent atoms in complex biochemical liquids. It improves and extends previous extrapolation schemes by including additionally new techniques such as an exponential scaling transformation of coordinate space with weights complemented by a dynamically adjusted balancing between the least square minimization of force deviations and the norm of expansion coefficients in the approximation. The expensive mean potential forces are treated in terms of the 3D-RISM-KH molecular theory of solvation (three-dimensional reference interaction site model with the Kovalenko-Hirata closure). During the dynamics they are calculated only after every long enough (outer) time interval, i.e., quite rarely to reduce the computational costs. At much shorter (inner) time steps, these forces are extrapolated on the basis of their outer values. The equations of motion are then solved using a multiple time step integration within an optimized isokinetic Nos\'e-Hoover chain thermostat. The new approach is applied to molecular dynamics simulations of various systems consisting of solvated organic and biomolecules of different complexity. Namely, we consider hydrated alanine dipeptide, asphaltene in toluene solvent, miniprotein 1L2Y and protein G in aqueous solution. It is shown that in all these cases, the enhanced extrapolation provides much better accuracy of the solvation force approximation than the existing approaches. As a result, it can be used with much larger outer time steps, leading to a significant speedup of the simulations.

\end{abstract}

\maketitle

\section{Introduction}

Molecular dynamics (MD) is one of the most important methods in studying various properties of different systems. Being originally designed sixty years ago \cite{Alder} to simulate elastic collisions between hard spheres, it was further extended to investigate simple liquids with more realistic potentials of interactions between particles. \cite{Allen, Frenkel, Leimkuhler, Tuckerman} During the last several decades, the method of MD has been developed to describe much more complicated systems, such as proteins in aqueous solution. \cite{McCammon, Brooks, Rojnuckarin, Duan, Hernandez, Karplus, Zhang, KarplusK, Adcock, Freddolino, Klepeis, Service, Shaw} However, prediction of the structure and functioning of proteins in computer simulations still remains a challenging task. \cite{Freddolinoa} The main problem is that the processes responsible for conformational and folding equilibria in these complex bioliquids take place on time scales ranging from microseconds up to minutes. \cite{Rojnuckarin, Duan, Hernandez, Karplus, Zhang, KarplusK} With the present capabilities of high-performance computers, the all-atom MD simulations of large proteins are limited, as a rule, to hundreds of nanoseconds. \cite{Genheden} This is, of course, insufficient to observe folding events even for simple proteins. Despite the development of massively parallel computing hardware including inexpensive graphics processing units (GPUs), it has remained infeasible to simulate the folding of atomistic proteins using conventional MD beyond the microsecond scale. \cite{Adhikari}

Various improvements to the conventional MD method have been proposed to obviate the problem with long time scales. Among them it is worth mentioning the replica exchange approach, hyperdynamics, implicit-solvent techniques, as well as the combination of MD with a molecular theory of solvation (many other approaches exist for increasing computational efficiency of atomistic MD simulations and for speeding up conformational sampling, for a review see, e.g., Refs.~\cite{DMZuckerman, Laio, Tuszynski}). In the replica exchange approach, \cite{Sugita, Pitera, Okur, Paschek, Kannan, Day, Mitsutake} a large number of short runs are carried out in parallel at different temperatures. During such runs, after certain time intervals the spatial configurations of macromolecules are periodically exchanged with a Metropolis rate (like in the Monte-Carlo method). Due to the presence of high-temperature replicas this allows to increase the probability for overcoming local minima separated by high-energy barriers inherent in the protein energy landscape. As a result, the necessary simulation length corresponding to each replica will be much shorter than the real folding time. However, the whole simulations must cover a wide temperature range with levels spaced closely enough to enable exchanges with high acceptance ratios. This significantly increases the total computational expenses.

In the hyperdynamics, \cite{Voter, Wereszczynski, Sinko, Pierce} the potential energy landscape is modified by raising energy minima that lower below a defined threshold level, while leaving those areas lying above the threshold unchanged. As a result, barriers separating adjacent energy basins are effectively reduced, providing the simulation access to conformational space that cannot be easily accessed in convenient MD simulations. Using the inherent power of GPUs, it was shown \cite{Pierce} that hyperdynamics simulations over several hundreds of nanoseconds are able to deal with conformational changes in proteins that typically occur on the millisecond time scale. However, the hyperdynamics approach requires the construction of biased potentials which should be equal to zero at transition states and positive in minima in order to accelerate the dynamics. Such a construction appears to be not trivial, especially for large proteins. Moreover, the long-time behavior obtained from the hyperdynamics can differ from the true dynamics due to the possible presence of transition-violating correlated events.

Another way to simplify the MD simulations of proteins is to replace all explicit atoms of solvent (water) molecules by a dielectric continuum with a predefined permittivity. Then we come to implicit-solvent methods which drastically reduce the number of particles to keep track of in the system and, thus, significantly speed up MD simulations. An extra effective acceleration comes from much faster sampling of the conformational space afforded by implicit solvent potentials which are much smoother in coordinate than the original interactions. In the context of hydration of biomolecules, the implicit-solvent methods can reproduce polar solvation forces with either the generalized Born \cite{Still, Onufriev, Onufrieva} or the Poisson-Boltzmann \cite{Antosiewicz} models, while nonpolar interactions are empirically accounted by the solvent accessible surface area model supplemented with additional volume and dispersion integral terms. \cite{Wagoner, Mongan} However, because of their empirical nature, such methods, involving a lot of adjustable parameters, lead to a less accurate description than the explicit-solvent approach. \cite{Pierce, Anandakrishnan} Moreover, they work well only for the hydration free energy but are not transferable to other solvents, cosolvents and solvent systems, in particular, to electrolyte solutions. In addition, the implicit-solvent methods cannot describe missing solvent size effects such as a desolvation barrier in protein aggregation and are inadequate to reproduce solvation of internal cavities such as narrow channels.

The above drawbacks of the implicit-solvent methods are absent in the 3D-RISM integral equation theory \cite{Chandler:1986:85:5971, Chandler:1986:85:5977, Beglov:1995:101:7821, Kovalenko:1998:290:237, Kovalenko:1999:110:10095, Kovalenko:2000:112:10391, Kovalenko:2000:112:10403, Kovalenko:2003:169, Hansen-McDonald:2006, Gusarov:2012:33:1478, Kovalenko:2013:85:159, Kovalenko:2015:22:575, Kafnn, Kobryng, Kafnm, KovGus} of molecular liquids (three-dimensional reference interaction site model) complemented with the Kovalenko-Hirata (KH) closure. \cite{Kovalenko:1999:110:10095, Kovalenko:2003:169, Kovalenko:2013:85:159} In the hybrid MD/3D-RISM-KH simulations, individual trajectories and dynamics of solvent molecules are contracted to quasiequilibrium 3D density distribution functions of their interaction atomic sites around the solute biomolecule in successive conformation snapshots. The time evolution of the biomolecule becomes quasidynamics steered with mean solvation forces obtained for each conformation of the biomolecule from the 3D-RISM-KH molecular theory of solvation. \cite{Miyata:2008:29:871, Luchko:2010:6:607, Omelyan:2013:39:25, Omelyan:2013:139:244106, Omelyann} The latter is derived from the first principles of statistical mechanics and uses explicitly the atomistic interaction potentials of the biomolecule and solvent molecules (force field). The 3D-RISM-KH mean solvation forces statistically averaged over the distributions of an infinite number of solvent molecules are thus added to the direct intramolecular interactions for integrating the equations of motion of atoms in the biomolecule. A chief advantage of such a hybrid approach is that slow processes in the system, such as reequilibration of solvent due to conformational changes of the solute biomolecule, distribution of ions and protein-ligand binding (which constitute a major challenge for conventional MD) are readily accounted for by 3D-RISM-KH mean solvation forces and excluded from the quasidynamics. This leads to a radical squeezing of time scales and, thus, to a substantial speedup of the simulations.

Pioneering MD/3D-RISM-KH simulations have been carried out by Miyata and Hirata \cite{Miyata:2008:29:871} for hydrated acetylacetone using the standard reference system propagator algorithm (RESPA) \cite{Tuckerman:1992:97:1990, Stuart:1996:105:1426, Kopf:1997:101:1} in the microcanonical ensemble to integrate the equations of motion. However, the maximal time steps were limited only to 5 fs because of resonance instabilities. \cite{Schlick:1997:26:181, Watanabe:1995:99:5680, Mandziuk:1995:237:525, Barth:1998:109:1633, Schlick:1998:140:1, Ma:2003:24:1951} The latter appear in conventional MD and hybrid MD/3D-RISM-KH simulations due to the multiple time step (MTS) interplay between strong intramolecular (solute-solute) and weak intermolecular (solute-solvent) forces. In conventional MD, the accuracy of MTS simulations can be increased by carrying out processed phase-space transformations. \cite{Omelyan:2008:78:026702, Omelyan:2009:131:104101} Employing these transformations within an energy-constrained scheme, it was demonstrated \cite{Omelyan:2011:135:114110} in MD simulations of water that outer time steps up to 16~fs are possible. But such steps cannot exceed the theoretical limiting value of 20~fs inherent in the microcanonical description. Moreover, in the MD/3D-RISM-KH simulations \cite{Miyata:2008:29:871} the integral equations were solved too frequently (every 5~fs), significantly slowing down the calculations. In order to damp the MTS instabilities, the MD/3D-RISM-KH approach has been extended \cite{Luchko:2010:6:607} to the canonical ensemble within the Langevin dynamics. \cite{Loncharich:1992:32:523, Barth:1998:109:1617} Introducing a method of solvation force extrapolation (SFE), it has been shown for hydrated alanine dipeptide that time steps up to 20~fs are acceptable. \cite{Luchko:2010:6:607} They, however, are still smaller than those available in conventional MD simulations by the isokinetic Nos\'e-Hoover chain RESPA (INR) integrator, for which outer time steps of 100~fs or even larger are possible. \cite{Minary:2004:93:150201, Abrams:2006:703:139, Minary:2003:118:2510, Omelyan:2011:135:234107, Omelyan:2012:8:6, Leimkuhlera, Margul, Chen}

Not so long ago, an optimized isokinetic Nos\'e-Hoover chain (OIN) canonical ensemble has been derived for more efficient elimination of MTS instabilities in MD simulations. \cite{Omelyan:2013:39:25} It improves the INR method \cite{Minary:2004:93:150201, Abrams:2006:703:139} and other canonical-isokinetic schemes \cite{Minary:2003:118:2510, Omelyan:2011:135:234107, Omelyan:2012:8:6, Leimkuhlera, Margul, Chen} by coupling each set of Nos\'e-Hoover chain thermostats to some optimal number of degrees of freedom in the system. Slightly modifying SFE of Ref.~\onlinecite{Luchko:2010:6:607}, the OIN integrator has been jointed with the MD/3D-RISM-KH approach. On an example of alanine dipeptide dissolved in water it has been proven \cite{Omelyan:2013:39:25} that the OIN ensemble is superior to the Langevin and INR schemes. In particular, large outer time steps of order of several hundred femtoseconds can be employed, providing a speedup up to 20 times with respect to conventional explicit-solvent MD. A method of advanced solvation force extrapolation (ASFE) in MD/3D-RISM-KH simulations has been developed, too. \cite{Omelyan:2013:139:244106} Here, a global non-Eckart-like rotation of atomic coordinates was utilized to minimize the distances between the biomolecule sites in different conformations at successive time steps. Then, extending the list of outer (reference) configurations, it has been shown that ASFE can provide a significantly better accuracy of the force evaluation than SFE. This has allowed to apply huge outer time steps up to tens of picoseconds without affecting equilibrium and conformational properties. As a result, the MD/OIN/ASFE/3D-RISM-KH simulations have accelerated by a factor of 100 to 500 compared to explicit solvent models. However, the applications were restricted to a relatively simple system of hydrated alanine dipeptide.

Recently, \cite{Omelyann} the ASFE approach was extensively modified and generalized to obtain nearly the same speedup for more complicated systems, including proteins. Rather than carrying out a rotational of the whole molecule, individual non-Eckart-like transformations were performed for each atom of the biomolecule. The individual scheme appreciably accelerated convergence of the extrapolated forces to their exact values with increasing the number of basic outer coordinates. Other techniques, such as an extension of the force-coordinate pair list to select the best subset and static balancing of the normal equations have been reconstructed, too. This resulted in a generalized SFE (GSFE) approach. It was demonstrated that GSFE can reach a high level of accuracy of the solvation force approximation at huge outer steps of order of 1 to 2~ps even for proteins. The MD/OIN/GSFE/3D-RISM-KH simulations provided a 50- to 1000-fold effective speedup of conformational sampling compared to conventional MD. The GSFE/3D-RISM-KH approach complemented by the OIN integrator has been implemented in the well-known and widely-used AMBER package of biomolecular simulation programs. \cite{Amber} Using this approach we have been able \cite{Omelyann} to fold the 1L2Y miniprotein from a fully extended state in about 60 ns of the 3D-RISM-KH quasidynamics for the first time, in contrast to an average physical folding time of 4-9 $\mu$s expected in conventional MD and observed in real experiment. \cite{Qiu, Snow} Note that so far, there have been no publications on folding this simplest protein by conventional MD despite the existence of highly specialized supercomputers, like Anton. \cite{Anton} The reason is that the $\mu$s scale is still practically unreachable in one MD run for most supercomputers, while the use of specialized hardware will lead to huge expenses of time and efforts.

In the present study, we go beyond GSFE and further improve the extrapolation strategy by additionally advancing to two new techniques such as an exponential scaling transformation of coordinate space with weights and a dynamical balancing between the minimization of force deviations and the norm of expansion coefficients arising during the approximation. The first one is aimed at better linearization and smoothing of solvation forces, resulting in an overall increase of the accuracy of the extrapolation. The second technique provides exact results in limits when the current spatial configuration appears to be close enough to any one of those containing in the reference list. This is contrary to the previous approximation schemes where the approximated values of solvent forces in such limits do not coincide with those related to the reference configurations. The new enhanced approach is applied to MD/OIN/3D-RISM-KH simulations of different solvated organic and biomolecular systems including proteins. It is shown that the enhanced extrapolation provides much better accuracy of the solvation force approximation than the existing approaches and can be used with much larger outer time steps, leading to a significant acceleration of the simulations.

\section{3D-RISM-KH theory of solvation}

\textit{Model.} --- Let us consider a solute macromolecule (protein) consisting of $M$ atoms dissolved in liquid composed of a large number of solvent molecules with $M'$ atomic sites. The potential energy of such a system can be cast in the form \cite{Amber}
\begin{align}
U({\bf r}_1, {\bf r}_2, \ldots, {\bf r}_{M+M'}) = & \!\sum_{i,j}^{\rm nonbonds}\! \bigg(\frac{a_{ij}}{r_{ij}^{12}}- \frac{b_{ij}}{r_{ij}^6}\bigg) + \!\sum_{i,j}^{\rm electrostatic} \frac{q_i q_j}{r_{ij}} + \!\sum_i^{\rm bonds} k_{b,i} (b_i-b_{i,0})^2 \nonumber \\ & + \!\sum_i^{\rm angles}\! k_{\theta,i} (\theta_i-\theta_{i,0})^2 + \!\sum_i^{\rm dihedrals}\! k_{\phi,i} \big(1+\cos[n_i \phi_i-\varphi_i]\big) \, ,
\label{eq:1}
\end{align}
where the summations are performed over all $M$ solute and $M'$ solvent atoms with $i \ne j$ if $i$ and $j$ belong to the same atomic type. In the rhs of Eq.~(1), the first term denotes the van der Waals interactions modeled by a Lennard-Jones function. The second one represents the electrostatic potential between atoms $i$ and $j$ with separation $r_{ij}=|{\bf r}_i-{\bf r}_j|$ and is given by Coulomb interactions. The third term is the potential between two chemically-bound atoms, modeled as a simple harmonic potential. The fourth contribution represents a bond-angle dependence involving three atoms and is also modeled by a harmonic potential. Finally, the fifth term relates to a dihedral angle (torsion) potential which is periodic and depends on four atom coordinates. Eq.~(1) requires the following type specific parameters: $a_{ij}$ (repulsion), $b_{ij}$ (attraction), $q_i$ (charge), $k_{b,i}$ (bond strength), $b_i$ (bond length), $b_{i,0}$ (equilibrium bond length), $k_{\theta,i}$ (angle strength), $\theta_i$ (bond angle), $\theta_{i,0}$ (equilibrium bond angle), $k_{\phi,i}$ (barrier for rotation), $\phi_i$ (dihedral angle), $n_i$ (number of maxima) and $\varphi_i$ (angular offset). They are fixed for a certain model and obtained from so-called force fields based on quantum mechanical calculations and experimental data. \cite{Duan:2003:24:1999, Simmerling:2002:124:11258, Wang:2004:25:1157}

\textit{Mean-potential forces.} --- Conventional MD simulations deal with instantaneous forces $-\partial U/\partial {\bf r}_i$ acting on all the particles ($i=1,2,\ldots,M+M')$ of the solute-solvent system under investigation. Here, the number of solvent molecules should be much larger (ideally infinite) than that of solute ones, i.e. $M' \gg M$, to have a good statistics and neglect the finite-size effects. In particular, for a biomolecule with $M \sim 10^3-10^4$ atoms such a number has to be of order of $M' \sim 10^5-10^6$ or more. This appreciably complicates the simulations because a vast majority of the computational costs is spent on the evaluation of intermolecular solute-solvent and solvent-solvent potentials. Note that in common practice, the concentration of solute macromolecules is small and the interactions between them are neglected (infinite dilution limit). Since we are interested exclusively in the study of conformational and folding behavior of a solute biomolecule, there is no sense in considering explicitly the dynamics of a huge number of solvent atoms. In view of this the main idea of an abbreviated description consists in the following. Firstly, we divide the full potential energy (1) into solute-solute $\mathcal{U}$ and solute-solvent $\mathfrak{U}$ contributions,
\begin{equation}
U({\bf r}_1, {\bf r}_2, \ldots, {\bf r}_{M+M'}) = \mathcal{U}({\bf r}_1, {\bf r}_2, \ldots {\bf r}_M) + \mathfrak{U}({\bf r}_1, {\bf r}_2, \ldots, {\bf r}_{M+M'}) \, ,
\label{eq:2}
\end{equation}
where ${\bf r}_i$ denotes the position of atom $i$ and the last term in the rhs of Eq.~(2) includes also solvent-solvent interactions. Secondly, instead to deal with the instantaneous solute-solvent and solvent-solvent potentials we can replace them by their mean-potential counterpart, \cite{Kirkwood,McQuarrie}
\begin{equation}
\mathfrak{\overline U}({\bf r}_1, {\bf r}_2, \ldots, {\bf r}_M) = \frac{{\displaystyle \int} \mathfrak{U} \ e^{-\frac{U}{k_{\rm B} T}}\, d {\bf r}_{M+1} d {\bf r}_{M+2}
\ldots d {\bf r}_{M+M'}}{{\displaystyle \int} e^{-\frac{U}{k_{\rm B} T}}
\, d {\bf r}_{M+1} d {\bf r}_{M+2} \ldots d {\bf r}_{M+M'}} \, ,
\label{mpf}
\end{equation}
where $k_{\rm B}$ denotes the Boltzmann constant and $T$ is the temperature of the system. The mean potential in Eq.~(3) is obtained by statistically averaging $\mathfrak{U}$ over all possible configurations of all $M'$ solvent atoms at a given (currently fixed) conformation $\{ {\bf r}_1, {\bf r}_2, \ldots {\bf r}_M \}$ of the solute macromolecule.

Note that $\mathfrak{\overline U}$ depends only on $M$ solute atomic positions in contrast to the original potential $\mathfrak{U}$ which is a function of $M+M'$ coordinates of all atoms in the system. This significantly simplifies the consideration because $M \ll M'$. Taking into account Eqs.~(2) and (3), the original dynamics of the system can be reduced to a quasidynamics of the macromolecule in the presence of mean-potential forces $\bm{\mathfrak{f}}_i = -\partial \mathfrak{\overline U}/\partial {\bf r}_i$, where $i=1,2,\dots,M$. The total force acting on atom $i$ of the macromolecule will then be equal to ${\bf f}_i = \bm{f}_i + \bm{\mathfrak{f}}_i$, where $\bm{f}_i = -\partial \mathcal{U}/\partial {\bf r}_i$ are the instantaneous forces due to the interactions between solute atoms inside the macromolecule. Another chief advantage of the reduced description is that the replacement of $\mathfrak{U}$ by $\mathfrak{\overline U}$ appreciably enhances sampling of protein conformational space. This follows from the fact that averaging out solvent degrees of freedom to mean-potential forces eliminates an astronomical number ($M' \gg M$) of local minima in the energy landscape of $\mathfrak{U}$ arising from local solvation structure fluctuations. As a result, we come to the $\mathfrak{\overline U}$-quasidynamics in which the most slow processes (such as reequilibration of solvent due to conformational changes of the solute macromolecule) are excluded, leading to a substantial squeezing of the time scale.

\textit{3D-RISM-KH approach.} --- Despite this time-scale squeezing, the quasidynamics requires an explicit form of the mean-potential forces. Eq.~(3) cannot be applied to practical calculations because it leads to an undoable $M'$-dimensional integration ($M' \sim 10^5-10^6$). However, such an integration can be obviated by using the 3D-RISM-KH molecular theory of solvation. \cite{Kovalenko:1998:290:237, Kovalenko:1999:110:10095, Kovalenko:2000:112:10391, Kovalenko:2000:112:10403, Kovalenko:2003:169, Hansen-McDonald:2006, Gusarov:2012:33:1478, Kovalenko:2013:85:159, Kovalenko:2015:22:575, Kafnn, Kobryng, Kafnm, KovGus} The 3D-RISM integral equations are derived from the 6D-Ornstein-Zernike relation \cite{Hansen-McDonald:2006} by partial averaging over orientations of solvent molecules around their interaction sites $\alpha$ to contract orientational degrees of freedom of the system. The result is \cite{Kovalenko:1999:110:10095, Kovalenko:2003:169}
\begin{equation}
g^{\rm uv}_\alpha({\bf r}) = 1 + \sum_\beta \int d{\bf r}^\prime c^{\rm uv}_\beta({\bf r}-{\bf r}^\prime) \chi^{\rm vv}_{\beta\alpha}(r^\prime) \, ,
\label{eq:3D-RISM}
\end{equation}
where $g^{\rm uv}_\alpha({\bf r})$ and $c^{\rm uv}_\beta({\bf r})$ are the 3D distribution and direct correlation functions for solvent atoms of types $\alpha$ and $\beta$, respectively, while $\chi^{\rm vv}_{\alpha\beta}(r)$ denotes the site-site susceptibility of the solvent (the superscripts ``u'' and ``v'' stand for sol\emph{u}te and sol\emph{v}ent species). Function $g^{\rm uv}_\alpha({\bf r})$ describes the 3D distribution of interaction site $\alpha$ of solvent molecules at position ${\bf r}$ around the solute macromolecule. The spatial convolution in Eq.~(4) can be calculated by exploiting the 3D fast Fourier transform on a supercell for the short-range parts of the correlations, while the long-range electrostatic asymptotics of the correlation functions is separated out and treated analytically. \cite{Kovalenko:2000:112:10391, Kovalenko:2003:169, Kaminski:2010:114:6082, Perkyns:2010:132:064106, Genheden:2010:114:8505, Gusarov:2012:JCC} The radially dependent susceptibility function $\chi^{\rm vv}_{\alpha\beta}(r) = \varsigma^{\rm vv}_{\alpha\beta}(r)
+ (g^{\rm vv}_{\alpha\beta}(r)-1) \rho^{\rm v}_\alpha$ is calculated in advance, where $\varsigma^{\rm vv}_{\alpha\beta}(r)$ denotes the intramolecular correlation function specifying the geometry of solvent molecules. The radial distribution function $g^{\rm vv}_{\alpha\beta}(r)$ in pure solvent with site number density $\rho^{\rm v}_\alpha$ is obtained from the dielectrically consistent version of the 1D-RISM formalism. \cite{Perkyns:1992:97:7656}

The set of integral equations (4) to be solved uniquely with respect to $g^{\rm uv}_\alpha({\bf r})$ and $c^{\rm uv}_\alpha({\bf r})$ must be complemented by a closure relation. The exact relation can be expressed \cite{Hansen-McDonald:2006} as a series of multiple integrals of the total correlation function $\zeta^{\rm uv}_\alpha({\bf r}) = g^{\rm uv}_\alpha({\bf r})-1$. However, being computationally intractable, it is replaced in practice with amenable approximations
which should analytically ensure asymptotics of the correlation functions and features of the solvation structure and thermodynamics to properly represent the solvation physics. The KH closure by Kovalenko and Hirata \cite{Kovalenko:1999:110:10095, Kovalenko:2003:169, Kovalenko:2013:85:159} just satisfies these criteria. It reads
\begin{equation}
g^{\rm uv}_\alpha({\bf r}) = \left\{
\begin{array}{ccc}
\exp\big[{\xi^{\rm uv}_\alpha({\bf r})}\big]
& \textrm{for} & \xi^{\rm uv}_\alpha({\bf r}) \le 0
\\ [12pt]
1 + \xi^{\rm uv}_\alpha({\bf r})
& \textrm{for} & \xi^{\rm uv}_\alpha({\bf r}) > 0
\end{array} \right. \, ,
\label{eq:3D-KH}
\end{equation}
where $\xi^{\rm uv}_\alpha({\bf r}) = - u^{\rm uv}_\alpha({\bf r})/(k_{\rm B}T) + \zeta^{\rm uv}_\alpha({\bf r}) - c^{\rm uv}_\alpha({\bf r})$ are the indirect functions and $u^{\rm uv}_\alpha({\bf r}) = \sum_i u^{\rm uv}_{i\alpha}(|{\bf r}-{\bf r}_i|)$. The interaction atom-atom potential $u^{\rm uv}_{i\alpha}(|{\bf r}-{\bf r}_i|)$ between solvent atom of type $\alpha$ located in ${\bf r}$ and solute site $i$ located at ${\bf r}_i$ are explicitly determined according to a force field (see Eq.~(1)). Having the correlation functions, the solvation free energy of the solute macromolecule can readily be derived in a closed analytical form as \cite{Kovalenko:1999:110:10095, Kovalenko:2003:169, Kovalenko:2013:85:159} \begin{align}
\mu_{\rm solv} = k_{\rm B}T \sum_\alpha \rho^{\rm v}_\alpha
\int d{\bf r} \left( \frac12 \big(\zeta^{\rm uv}_\alpha({\bf r})\big)^2
\Theta\big(-\zeta^{\rm uv}_\alpha({\bf r})\big) - c^{\rm uv}_\alpha({\bf r}) - \frac12 \zeta^{\rm uv}_\alpha({\bf r})
c^{\rm uv}_\alpha({\bf r}) \right) ,
\label{eq:xMu-3D-KH}
\end{align}
where $\Theta$ denotes the Heaviside function. Then the solvation forces $\bm{\mathfrak{f}}_i$, acting on each atom $i$ of the solute macromolecule and representing the mean-potential ones for our system, are calculated by spatially differentiating $\mu_{\rm solv}$ with respect to ${\bf r}_i$, where $i=1,2,\ldots,M$. In view of Eq.~(6) this yields \cite{Miyata:2008:29:871,Luchko:2010:6:607}
\begin{equation}
\bm{\mathfrak{f}}_i = - \frac{\partial \mathfrak{\overline U}}{\partial {\bf r}_i} = -\frac{\partial
\mu_{\rm solv}}{\partial {\bf r}_i} = \sum_\alpha
\rho^{\rm v}_\alpha \int d{\bf r} g^{\rm uv}_\alpha({\bf r})
\frac{\partial u^{\rm uv}_{i \alpha}({\bf r}-
{\bf r}_i)}{\partial {\bf r}_i} \, .
\label{eq:forces}
\end{equation}

The 3D-RISM integral equation (4) with the KH closure (5) are solved numerically by iterations using the modified algorithm of direct inversion in the iterative subspace (MDIIS). \cite{Kovalenko:2000:112:10391, Kovalenko:2000:112:10403, Kovalenko:2003:169, Gusarov:2012:33:1478, Kovalenko:1999:20:928, Kovalenko:1999:103:7942} It accelerates convergence of integral equations by optimizing each iterative solution in a Krylov subspace of typically last 10-20 successive iterations and then making the next iterative guess by mixing the optimized solution with the approximated optimized residual. Memory and CPU load in the MDIIS numerical solver are decreased by up to an order of magnitude using the core-shell-asymptotics treatment of solvation shells. \cite{Gusarov:2012:33:1478} The computational expenses can further be significantly reduced with several strategies, including a high-quality initial guess for the 3D direct correlation functions $c^{\rm uv}_\alpha(\bf{r})$; pre- and post-processing of the 3D solute-solvent potentials $u^{\rm uv}_\alpha({\bf r})$, the long-range asymptotics of the 3D correlation functions $c^{\rm uv}_\alpha(\bf{r})$ and $g^{\rm uv}_\alpha(\bf{r})$ as well as solvation forces; several cutoff schemes and an adaptive solvation box. \cite{Luchko:2010:6:607} Additional speedup can be reached by using a multigrid version of the MDIIS algorithm. \cite{Sergiievskyi}

\section{Enhanced solvation force extrapolation (ESFE)}

\textit{Exponential scaling linearization with weights.} --- Though the above accelerated convergence of the 3D-RISM-KH integral equations, the calculation of solvation forces (7) requires, nevertheless, much larger computational efforts than that of intramolecular interactions $\bm{f}_i = -\partial \mathcal{U}/\partial {\bf r}_i$. Thus, the idea is to converge the expensive integral equations only after every long enough (outer) time interval during the quasidynamics. At much shorter (inner) time steps, these forces can be approximated using a fast extrapolation. Indeed, contrary to the instantaneous solute-solvent interactions $-\partial \mathfrak{U}/\partial {\bf r}_i$ evaluating in conventional MD, the solvation forces $\bm{\mathfrak{f}}_i = -\partial \mathfrak{\overline U}/\partial {\bf r}_i$ are relatively smooth. This means that the latter vary with changing time and coordinates much slower than the former. The reason is that the solvation forces are obtained at a given conformational state $\{ {\bf r}_1, \ldots, {\bf r}_M \}$ of the solute atoms by statistically averaging over all possible equilibrium configurations of the solvent molecules. As a result, the repulsive cores and other strong components (see Eq.~(1)), existing in instantaneous solute-solvent interactions $-\partial \mathfrak{U}/\partial {\bf r}_i$, will be merely absent after the averaging in $\bm{\mathfrak{f}}_i=-\partial \mathfrak{\overline U}/\partial {\bf r}_i$. Because of this smoothness, the 3D-RISM-KH forces allow to be extrapolated, so that the integral equations can be handled less frequently, increasing the efficiency of the MD simulations.

According to Eq.~(7), the solvation force $\bm{\mathfrak{f}}_i$ acting on a given atom $i$ (where $i=1,2,\ldots,M$) of the macromolecule actually depends on $M-1$ vector coordinates ${\bf r}_{ij}={\bf r}_i-{\bf r}_j$, where $j \ne i$, which define the relative positions of all neighbouring atoms around reference site $i$, i.e.,
\begin{equation}
\bm{\mathfrak{f}}_i \equiv \bm{\mathfrak{f}}_i({\bf r}_1, {\bf r}_2, \ldots, {\bf r}_M) \equiv \bm{\mathfrak{f}}_i({\bf r}_{i1}, \ldots, {\bf r}_{ii-1}, {\bf r}_{ii+1}, \ldots, {\bf r}_M) \equiv \bm{\mathfrak{f}}_i(\{{\bf r}_{ij}\}) \, .
\end{equation}
This follows from the translational invariance of solvation interactions when the total system (solute plus solvent) is arbitrarily shifted as a whole. The first important step of our new enhanced approach is to perform such a scaling transformation of ${\bf r}_{ij}$ to new vectors $\bm{\varrho}_{ij}$ to obtain the most linear dependence of $\bm{\mathfrak{f}}_i$ on $\bm{\varrho}_{ij}$. Then we will be entitled to apply a linear extrapolation of $\bm{\varrho}_{ij}$, automatically providing the approximation of $\bm{\mathfrak{f}}_i$ with minimal uncertainties (see below). Remember that we deal with a $(M-1)$-multdimensional ($M \gg 1$) function $\bm{\mathfrak{f}}_i(\{{\bf r}_{ij}\})$ in which the explicit forms of the coordinate dependencies are unknown (the solvation forces are calculated numerically and cannot be expressed analytically). Because of this a full linearization is impossible in our case, but we can use general properties of $\bm{\mathfrak{f}}_i(\{{\bf r}_{ij}\})$ to linearize it in part. They are: (i) effective magnitude of $\bm{\mathfrak{f}}_i$ decays with increasing interatomic separations $r_{ij}=|{\bf r}_{ij}|$ at their large enough values, leading to the limiting behaviour $\lim_{\{r_{ij\}} \to \infty} \bm{\mathfrak{f}}_i(\{{\bf r}_{ij}\})=0$, (ii) at a given $r_{ij} \ne 0$, the contribution to solvation force acting on atom $i$ caused by atom $j$ ($j \ne i$) depends on type of the latter and is negligible at $r_{ij}=0$, and (iii) the total solvation force acting on the macromolecule as a whole is equal to zero, i.e., $\sum_i \bm{\mathfrak{f}}_i = 0$.

Many analytical expressions can be involved to model the above properties. The most simple choice is an exponential scaling of vectors ${\bf r}_{ij}$ without changing their directions,
\begin{equation}
\bm{\varrho}_{ij} = w(r_{ij}) {\bf r}_{ij} \, , \ \ \ \ \ \ \
w(r_{ij}) = w_{ij} \exp(-\eta_{ij} r_{ij}) \, ,
\end{equation}
where $w(r_{ij})$ denote the transformation functions, $\eta_{ij} = \eta_{ji} \ge 0$ are the scaling parameters and $w_{ij} = w_i w_j > 0$ stand for the weights obeying the normalization $\sum_i w_i^2 = M$. This scaling tries to reproduce feature (i) and exactly provides the true zeroth limiting behaviour since $\lim_{r_{ij} \to \infty} w(r_{ij}) = 0$. The weights try to take into account the fact (ii) that the influence of atoms on $\bm{\mathfrak{f}}_i(\{{\bf r}_{ij}\})$ can depend on their type. In addition, $\bm{\varrho}_{ij} = 0$ at $r_{ij}=0$ by construction. Moreover, $\sum_{ij} \bm{\varrho}_{ij} = 0$ as this is required by property (iii) since $\sum_{ij} {\bf r}_{ij} = 0$. Indeed, $w(r_{ij})$ depends only on magnitude of vector ${\bf r}_{ij}$ but not on its direction, while $\eta_{ij}$ together with $w_{ij}$ are symmetrical with respect to the substitution $i \leftrightarrow j$. For illustration of the scaling transformation, consider a virtual system of two ($M=2$) particles in one-dimensional space $-\infty < x_{1,2} < \infty$ influencing one on another through the forces $f_{1,2}(x) = \pm x \exp(-\eta |x|)$, where $x=x_1-x_2$, satisfying all of the above three properties. Then making the transformation $X = x \exp(-\eta |x|)$ fully linearizes this force in the new variable, i.e., $f_{1,2}(X) = \pm X$.

Apart from the unform distribution $w_i \equiv 1$, there are three additional variants to build the weights, namely,
\begin{equation}
w_i = \sqrt{\frac{\langle \bm{\mathfrak{f}}_i^2 \rangle}{\frac1M \sum_{j=1}^M \langle \bm{\mathfrak{f}}_j^2 \rangle}}, \ \ \ \ \
w_i = \frac{|q_i|}{\left(\frac1M \sum_{j=1}^M q_j^2\right)^{1/2}} , \ \ \ \ \
w_i = \frac{m_i}{\left(\frac1M \sum_{j=1}^M m_j^2\right)^{1/2}}
\end{equation}
corresponding to the averaged-force, charge, and mass weighting, respectively, where $q_i$ is the charge and $m_i$ is the mass of atom $i$. Note that in the force scheme, the weights are not constant and change during the simulations, where $\langle \ \rangle$ denotes the averaging over the produced atomic trajectories. The truncation $w(r_{ij}) \approx 0$ at large separations $r_{ij} > r_{{\rm c},ij}$ can be applied to reduce the computational costs, where $r_{{\rm c},ij} = r_{\rm c} + \ln w_j/\eta_{ij}$ and $r_{\rm c}$ is the fixed truncation radius. The concrete variant of the weighting scheme as well as the values of $\eta_{ij}$ and $r_{\rm c}$ are chosen in such a way to provide the best effective overall linearization.

\textit{Individual rotation transformations.} --- Let $\bm{\mathfrak{f}}_{i,k}$ be the solvation forces acting on solute sites $i=1,2,\ldots,M$ at $N$ previous outer time steps $k=1,2,\ldots,N$ for which the 3D-RISM-KH integral equations are converged. The relative atomic positions at these steps will be denoted by ${\bf r}_{ij,k}$. The forces $\bm{\mathfrak{f}}_{i,k}$ and positions ${\bf r}_{ij,k}$ for a given $i$ are ordered in such a way that larger numbers of $k$ correspond to earlier moments $t_k$ of time, i.e., $t_N < t_{N-1} < \ldots < t_2 < t_1$. The next outer moment is denoted by $t_0 > t_1$. Let ${\bf r}_{ij}(t)$ be the current relative coordinates at some inner time point $t$ belonging to the interval $]t_1, t_0[$. The total number of these points is equal to $P=(t_0 - t_1)/\Delta t = h/\Delta t \gg 1$, where $h = t_{k-1}-t_k$ and $\Delta t$ are the outer and inner time steps, respectively ($h \gg \Delta t$).

The second main idea of the new approach is to find such local rotational transformations ${\bf R}_{ij} = {\bf S}_i \bm{\varrho}_{ij} = {\bf S}_i w(r_{ij}) {\bf r}_{ij} \equiv w(r_{ij}) {\bf S}_i {\bf r}_{ij}$ of the scaled positions $\bm{\varrho}_{ij} = w(r_{ij}) {\bf r}_{ij}$ for each atom $i=1,2,\ldots,M$ (where $j=1,\ldots,i-1,i+1,\ldots,M$) that provide the most smooth behavior of the solvation forces $\bm{\mathfrak{F}}_i = \bm{\mathfrak{f}}_i(\{{\bf R}_{ij}\})$ in the transformed coordinates. For the discrete set ($k=1,2,\ldots,N$) of the basic coordinate knots ${\bf r}_{i,k}$, the desired transformation ${\bf R}_{ij,k}={\bf S}_{i,k} \bm{\varrho}_{ij,k} = w(r_{ij,k}) {\bf r}_{ij,k}$ with ${\bf r}_{ij,k}={\bf r}_{i,k}-{\bf r}_{j,k}$ can be determined by minimizing the distances between all the transformed outer coordinates ${\bf R}_{ij,k}$ and some origin point $\bm{\varrho}_{ij}^\ast = w(r_{ij}^\ast) {\bf r}_{ij}^\ast$ (where ${\bf S} \equiv {\bf I}$) lying in the extrapolating region as
\begin{equation}
\frac{1}{M_i} \sum_{j=1}^M { }^{'}
\Theta(r_{ij}^\ast-r_{{\rm c},j})
\Big( w(r_{ij,k}) {\bf S}_{i,k} {\bf r}_{ij,k} -
w(r_{ij}^\ast) {\bf r}_{ij}^\ast \Big)^2 = \min
\label{tcm}
\end{equation}
for each given $i=1,2,\ldots,M$ and $k=1,2,\ldots,N$. Here $M_i = \sum_{j=1}^M\!{ }^{'} \Theta(r_{ij}^\ast-r_{{\rm c},j}) w(r_{ij})$ is the effective number of neighbours and $\sum'$ stands for $j \ne i$. The current inner coordinate ${\bf r}_{ij}(t)$ should also be transformed analogously by ${\bf R}_{ij}(t) = {\bf S}_i(t) \bm{\varrho}_{ij} = w(r_{ij}) {\bf S}_i(t) {\bf r}_{ij}(t)$, where ${\bf S}_i(t)$ is found from the minimization
\begin{equation}
\frac{1}{M_i} \sum_{j=1}^M { }^{'}
\Theta(r_{ij}^\ast-r_{{\rm c},j})
\Big( w(r_{ij}) {\bf S}_i(t) {\bf r}_{ij}(t) -
w(r_{ij}^\ast) {\bf r}_{ij}^\ast \Big)^2 = \min \, .
\label{tcmc}
 \end{equation}
Note that ${\bf S}_{i,k} w(r_{ij,k}) {\bf r}_{ij,k} = w(r_{ij,k}) {\bf S}_{i,k} {\bf r}_{ij,k}$ and ${\bf S}_i(t) w(r_{ij}) {\bf r}_{ij}(t) = w(r_{ij}) {\bf S}_i(t) {\bf r}_{ij}(t)$ because the rotational matrices change only directions of vectors but not their lengths, i.e., ${\bf S}_{i,k} r_{ij,k} = r_{ij,k}$ and ${\bf S}_i(t) r_{ij}(t) = r_{ij}(t)$.

Any choice for the reference point ${\bf r}_{ij}^\ast = {\bf r}_{ij}(t^\ast)$ with $t_1 \le t^\ast \le t \le t_0$ can be in principle acceptable, where $t$ is the current inner time and $t_1$ is the most recent point from the basic outer steps. However, the limiting values $t^\ast=t_1$ and $t^\ast=t$ are not recommended in the context of efficiency. Note that in Eq.~(11) we should carry out the transformation for each $k=1,2,\ldots,N$ (and $i=1,2,\ldots,M$) whenever ${\bf r}_{ij}^\ast$ is changed, i.e., up $N M P$ times if $t^\ast=t$, increasing the costs, but only $N M$ times at $t^\ast=t_1$. In the latter case, however, the origin ${\bf r}_{ij}^\ast$ being equal to ${\bf r}_{ij,1}$ may appear to be too far from the current point ${\bf r}_{ij}(t)$ when the size of the outer time step $h=t_0-t_1$ is large, lowering the accuracy. Thus, an optimal choice is when the origin ${\bf r}_{ij}^\ast$ of the transformation is updated after every $1 \ll p \ll P$ inner time step during the outer interval $(t_0-t_1)$. This provides a good accuracy at reasonable computational costs and will be referred to as a frequency reuse regime. Note also that due to the presence of the weight scaling function $w(r_{ij})$, the $j$-neighbours with larger interatomic distances ($r_{ij} > 1/\sqrt{\eta_{ij}}$) give smaller contributions to Eqs.~(11) and (12). This is quite natural because the mean solvation forces $\bm{\mathfrak{f}}_i$ decrease in mean with increasing $r_{ij}$ at large separations. At long enough $r_{ij} > r_{{\rm c},j}$ the correlations between ${\bf r}_{ij}$ and $\bm{\mathfrak{f}}_i$ are diminished and we put $w(r_{ij})=0$ in this range. Such a truncation concerns only the scaling transformation (9) but is not applied when calculating the actual solvation forces $\bm{\mathfrak{f}}_{i,k}$.

It should be emphasized that the rotational superpositions (11) and (12) are carried out individually ($i=1,2,\ldots,M$) for each atom of the macromolecule. They effectively take into account local rotations of the solute molecule, which can be large due to the interactions with the solvent. As a result, the changes of solvation forces $\bm{\mathfrak{f}}_i(\{{\bf r}_{ij}\})$ caused by such rotations will be merely excluded in the transformed coordinate space $\{{\bf R}_{ij}\}$ leading to the most smooth behaviour of $\bm{\mathfrak{F}}_i = \bm{\mathfrak{f}}_i(\{{\bf R}_{ij}\})$. For example, in the case of rotating rigid segments constituting the macromolecule, the transformed forces $\bm{\mathfrak{F}}_i$ will be constant at all and thus can be extrapolated exactly. These forces will be changed not so much even for flexible segments, since the magnitudes of the atomic vibrational oscillations are small. Note also that Eqs.~(11) and (12) look somewhat similar to those of Eckart or Eckart-like approaches used to separate translational, angular and internal motions of macromolecules. \cite{Eckart:1935:47:552, Louck:1976:48:69, Janezic:2005:122:174101, Praprotnik:2005:122:174102, Praprotnik:2005:122:174103, Kneller:2008:128:194101, Omelyan:2012:85:026706} Our non-Eckart superposition scheme differs in several aspects from the original Eckart method. \cite{Kneller:2008:128:194101} It is modified by weight scaling transformations and applied individually for each reference atom of the solute macromolecule. This results in local reorientations of atomic groups instead in a rotation of the molecule as a whole. Moreover, the newly introduced non-Eckart scheme is aimed at optimizing the performance of MD simulations rather than only at analyzing simulations or experimental data. \cite{Kneller:2008:128:194101, Coutsias:2004:25:1849, Liu:2009:31:1561, Chevrot:2011:114:6082}

The simplest way to obtain explicit expressions for the rotational matrix ${\bf S}_{i,k}$ and ${\bf S}_i(t)$ is to represent them in terms of the four components quaternion ${\bf q}=\{q_x,q_y,q_z,q_u\}$ as \cite{Omelyan:1999:22:213}
\begin{equation}
{\bf S} = \left(
\begin{array}{ccc}
q_x^2+q_y^2-q_z^2-q_u^2 & 2(q_yq_z-q_xq_u) & 2(q_xq_z+q_yq_u) \\ [4pt]
2(q_xq_u+q_yq_z) & q_x^2+q_z^2-q_y^2-q_u^2 & 2(q_zq_u-q_xq_y) \\ [4pt]
2(q_yq_u-q_xq_z) & 2(q_xq_y+q_zq_u) & q_x^2+q_u^2-q_y^2-q_z^2 \\ [4pt]
\end{array}
 \right)
\label{eq:qmt}
\end{equation}
with ${\bf q}^2 \equiv {\bf q}^+ {\bf q}=q_x^2+q_y^2+q_z^2+q_u^2=1$. Inserting Eq.~(13) into the superposition equations (11) and (12) yields
\begin{equation}
\frac12 {\bf q}^+ \Bigg( \frac{1}{M_i} \sum \limits_{j=1}^M{\!}^{'} \Theta(r_{ij}^\ast-r_{{\rm c},j}) \bm{\Phi}_{ij} \Bigg) {\bf q} - \frac12 \lambda
({\bf q}^+ {\bf q} - 1) = \min \, ,
\label{eq:tqm}
\end{equation}
where
\begin{equation}
\bm{\Phi}_{ij} \!=\!
\left(\!\!
\begin{array}{cc}
(\bm{\varrho}'_{ij}-\bm{\varrho}_{ij}^\ast)^2 &
2 (\bm{\varrho}'_{ij} \bm{\times} \bm{\varrho}_{ij}^\ast)^+ \\ [8pt]
2 (\bm{\varrho}'_{ij}\! \bm{\times} \bm{\varrho}_{ij}^\ast) \ \ &
{\bf I}(\bm{\varrho}'_{ij}\!+\bm{\varrho}_{ij}^\ast)^2\!-
2(\bm{\varrho}'_{ij} {\bm{\varrho}_{ij}^\ast}^+\! +
\bm{\varrho}_{ij}^\ast {\bm{\varrho}'_{ij}}^+) \\ [6pt]
\end{array}
\!\!\right)
\label{eq:qtm}
\end{equation}
are the symmetric $4\times4$ matrices, $\bm{\varrho}'_{ij}$ is equal either to $w(r_{ij,k}) {\bf r}_{ij,k}$ or $w(r_{ij}) {\bf r}_{ij}(t)$ for the cases ${\bf S}_{i,k}$ or ${\bf S}_i(t)$, respectively, $\lambda$ is the Lagrange multiplier, ${\bf I}$ is the identity $3 \times 3$ matrix, $\bm{\times}$ denotes the vector product, and $\bm{\varrho}_{ij}^\ast = w(r_{ij}^\ast) {\bf r}_{ij}^\ast$. Differentiating (14) with respect to all four components of ${\bf q}$ leads to the eigenvalue problem
\begin{equation}
\Bigg( \frac{1}{M_i} \sum \limits_{j=1}^M{\!}^{'} \Theta(r_{ij}^\ast-r_{{\rm c},j}) \bm{\Phi}_{ij} \Bigg) {\bf q} \equiv \bm{\Phi}_i {\bf q} = \lambda {\bf q} \, .
\label{eq:evp}
\end{equation}
Because the lhs of Eqs.~(11) and (12) are always greater or equal to zero, the matrix $\bm{\Phi}_i$ is positive semidefinite, leading to four eigenvectors ${\bf q}^\textrm{(1,2,3,4)}$ and the same number of nonnegative associated eigenvalues $\lambda^\textrm{(1,2,3,4)} \ge 0$ in Eq.~(16). The latter are sorted in the ascending order, so that $\lambda^\textrm{(1)}$ is the smallest eigenvalue. It coincides with the global minimum in Eqs.~(11), (12) and (14) since for any normalized eigenvectors ${\bf q}$ the following equality takes place: ${\bf q}^+ \bm{\Phi}_i {\bf q}=\lambda$. The normalized eigenvector ${\bf q}^\textrm{(1)}$ corresponding to the smallest eigenvalue $\lambda^\textrm{(1)} \equiv \lambda^\textrm{(1)}_{i,k}$ or $\lambda^\textrm{(1)}_i$ is thus the quaternion describing the desired transformation by the rotational matrix ${\bf S} \equiv {\bf S}_{i,k}$ or ${\bf S}_i(t)$ [see Eq.~(13)].

The next ideas of the enhanced approach are described below in the three successive subsections.

\textit{Least-square minimization of uncertainties.} --- The solvation forces $\bm{\mathfrak{F}}_i = \bm{\mathfrak{f}}_i(\{{\bf R}_{ij}\})$ can be represented in the transformed space as the power series of deviations of the current coordinate vectors ${\bf R}_{ij}$ from the origin values $\bm{\varrho}_{ij}^\ast$ for each $i=1,2,\ldots,M$ as
\begin{equation}
\bm{\mathfrak{F}}_i = \bm{\mathfrak{f}}_i(\{{\bf R}_{ij}\}) = \bm{\mathfrak{f}}_i(\{\bm{\varrho}_{ij}^\ast\}) +
\sum_{j=1}^M { }^{'} \frac{\partial \bm{\mathfrak{F}}_i}{\partial {\bf R}_{ij}}
\bigg|_{\bm{\varrho}_{ij}^\ast} \big({\bf R}_{ij} - \bm{\varrho}_{ij}^\ast) + \mathcal{O}[({\bf R}_{ij} - \bm{\varrho}_{ij}^\ast)^2] \, ,
\label{eq:fep}
\end{equation}
where $\bm{\Xi}_{ij}=\partial \bm{\mathfrak{F}}_i/\partial {\bf R}_{ij}$ is the Hessian ($3M \times 3M$) matrix. The second- and higher-order spatial inhomogeneities $\mathcal{O}[({\bf R}_{ij} - \bm{\varrho}_{ij}^\ast)^2]$ of $\bm{\mathfrak{f}}_i(\{{\bf R}_{ij}\})$ can be neglected because of the above scaling linearization and rotational transformations. Then from the form of Eq.~(17) it immediately follows that extrapolation of solvation forces $\bm{\mathfrak{f}}_i$ is reduced to an approximation of relative atomic coordinates ${\bf R}_{ij}$. Indeed, a better representation of ${\bf R}_{ij}$ automatically provides a more accurate extrapolation of interactions $\bm{\mathfrak{f}}_i$, because the latter are (linear) functions of only $\{{\bf R}_{ij}\}$ according to Eq.~(17).

Having the transformed coordinates at outer times steps, ${\bf R}_{ij,k} = w(r_{ij,k}) {\bf S}_{i,k} {\bf r}_{ij,k}$, and their current value ${\bf R}_{ij}(t) = w(r_{ij}(t)) {\bf S}_i(t) {\bf r}_{ij}(t)$ the latter can be extrapolated as follows. First, for each atom $i$, the actual neighbouring positions ${\bf R}_{ij}(t)$ are virtually approximated at a given inner point $t$ of the next outer time interval $]t_1, t_0[$ by a linear combination of their previous outer values as
\begin{equation}
{\bf \tilde R}_{ij}(t) =
\sum_{k=1}^{N} A_k^{(i)}(t) {\bf R}_{ij,k} \, .
\label{eq:ce}
\end{equation}
The expansion coefficients $A_k^{(i)}(t)$ in Eq.~(18) can then be obtained as the best representation of the solute neighbouring coordinates ${\bf R}_{ij}(t)$ at time $t$ in terms of their projections onto the basis of $N$ previous outer positions ${\bf R}_{ij,k}$ by minimizing the square norm of the difference between ${\bf R}_{ij}(t)$ and their approximated counterparts ${\bf \tilde R}_{ij}(t)$. Additionally imposing the normalizing condition on the coefficients and minimizing their square norm, i.e.,
\begin{equation}
\sum_{k=1}^N A_k^{(i)} = 1 \, , \ \ \ \ \ \ \ \sum_{k=1}^N {A_k^{(i)}}^2 = \min
\end{equation}
lead to the following least-square problem
\begin{equation}
\label{cemlt}
\frac{1}{M_i} \sum_{j=1}^M { }^{'} \Theta(r_{ij}^\ast-r_{{\rm c},j}) \left(
{\bf R}_{ij} - \sum_{k=1}^N A_k^{(i)} {\bf R}_{ij,k} \right)^{\!2}
\! + \, 2\Lambda_i \left(\sum_{k=1}^N A_k^{(i)}-1\right)
+ \, \varepsilon_i \mathcal{R}_i^2 \sum_{k=1}^N {A_k^{(i)}}^2 \!= \min
\end{equation}
for each $i=1,2,\ldots,M$. Here $\Lambda_i$ is the Lagrangian multiplier, while $\varepsilon_i >0$ and $\mathcal{R}_i^2 \ge 0$ denote the balance constants and dynamical functions, respectively, whose meaning and explicit forms will be presented below.

The forces $\bm{\mathfrak{F}}_i(t)$ at any inner time $t \in ]t_1, t_0[$ can be extrapolated on the basis of their outer values $\bm{\mathfrak{F}}_{i,k} = \bm{\mathfrak{f}}_i(\{{\bf R}_{ij,k}\})$ employing a linear expansion procedure which is quite similar to that for coordinates ${\bf \tilde R}_{ij}(t)$, namely,
\begin{equation}
\bm{\mathfrak{\tilde F}}_i(t) = \sum_{k=1}^{N} A_k^{(i)}(t) \bm{\mathfrak{F}}_{i,k} \, ,
\label{eq:fe}
\end{equation}
where $i=1,2,\ldots,M$ and the expansion coefficients $A_k^{(i)}(t)$ are the same as those in Eq.~(18). This is justified by the linearity of expansion (17). In such a way, the coordinate minimization (20) provides a minimization of deviations between the exact forces $\bm{\mathfrak{F}}_i(t)$ and their approximated values $\bm{\mathfrak{\tilde F}}_i(t)$ at each $i$, in the sense that $[\bm{\mathfrak{F}}_i(t) - \bm{\mathfrak{\tilde F}}_i(t)]^2 = [\sum'_j \bm{\Xi}_{ij} ({\bf R}_{ij} - {\bf \tilde R}_{ij})]^2 = \min$ according to Eqs.~(17) and (20). Note that the coordinate mapping is virtual meaning that ${\bf R}_{ij}(t)$ are never replaced by ${\bf \tilde R}_{ij}(t)$. It is necessary only to find coefficients for the real force approximation (21).

The transformed forces $\bm{\mathfrak{F}}_{i,k}$ can be obtained from original values $\bm{\mathfrak{f}}_{i,k}$ without direct recalculations ${\bf F}_{i,k}={\bf f}(\{ {\bf R}_{ij,k} \})$ by taking into account the following identities
\begin{equation}
\bm{\mathfrak{F}}_{i,k} = \bm{\mathfrak{f}}_i(\{{\bf R}_{ij,k}\}) = \bm{\mathfrak{f}}_i(\{{\bf S}_{i,k} \bm{\varrho}_{ij,k}\}) = {\bf S}_{i,k} \bm{\mathfrak{f}}_i(\{{\bf r}_{ij,k}\}) \equiv {\bf S}_{i,k} \bm{\mathfrak{f}}_{i,k} \, ,
\label{eq:tcon}
\end{equation}
\begin{equation}
\bm{\mathfrak{F}}_i(t) = \bm{\mathfrak{f}}_i(\{{\bf R}_{ij}(t)\}) = \bm{\mathfrak{f}}_i(\{{\bf S}_i \bm{\varrho}_{ij}(t)\}) = {\bf S}_i(t) \bm{\mathfrak{f}}_i(\{{\bf r}_{ij}(t)\}) \equiv {\bf S}_i(t) \bm{\mathfrak{f}}_i(t) \, .
\label{eq:tcona}
\end{equation}
They follow from the translational and orientational invariance of solvation forces. In particular, when a system is rotated as a whole, the solvation force vectors will also be rotated on the same angle around the some axis. In view of Eqs.~(22) and (23), no additional direct recalculations are needed, and the desired approximated forces in the usual coordinate space at each inner time point $t$ can be readily reproduced from Eq.~(21) using the inverse rotational transformation
\begin{equation}
\bm{\mathfrak{\tilde f}}_i(t) = {\bf S}_i^{-1}(t) \bm{\mathfrak{\tilde F}}_i(t) =
{\bf S}_i^{-1}(t) \sum_{k=1}^{N} A_k^{(i)}(t)
{\bf S}_{i,k} \bm{\mathfrak{f}}_{i,k} \, .
\label{eq:fets}
\end{equation}
The inverse matrix can easily be evaluated taking into account that the rotational transformation is orthonormal, i.e., ${\bf S}_i^{-1} = {\bf S}_i^+$, where ${\bf S}_i^+$ denotes the transposed matrix.

\textit{Normal equations with dynamical balancing.} --- There are several schemes to solve the least-square problem (20), including the QR-factorization and normal-equation method. \cite{Lawson:1974, Quintana:1999:20:1155} The latter is the most efficient way to find the coefficients $A_k^{(i)}$ for the force extrapolation (24). The normal representation can be obtained by differentiation of Eq.~(20) with respect to these coefficients and Lagrange multiplier $\Lambda_i$ for each $i=1,2,\ldots,M$. This leads to the following set of $N+1$ linear equations
\begin{equation}
\left(
\begin{array}{ccccc}
G_{11}^{\varepsilon (i)}  & G_{12}^{(i)}  &
\dots   &  G_{1N}^{(i)}   &  1    \\ [4pt]
G_{21}^{(i)}   & G_{22}^{\varepsilon (i)} &
\dots   &  G_{2N}^{(i)}   &  1    \\ [4pt]
\vdots  & \vdots & \vdots & \vdots & \vdots \\ [4pt]
G_{N1}^{(i)}   &  G_{N,2}^{(i)}  & \dots  &
G_{NN}^{\varepsilon (i)}  &  1 \\ [4pt]
1    &   1  &   \dots  &  1   &  0
\end{array}
\right) \left(
\begin{array}{c}
A_1^{(i)} \\ [4pt] A_2^{(i)} \\ [4pt] \vdots \\ [4pt]
A_N^{(i)} \\ [4pt] \Lambda_i
\end{array}
\right) = \left(
\begin{array}{c}
G_1^{(i)} \\ [4pt] G_2^{(i)} \\ [4pt] \vdots \\ [4pt]
G_N^{(i)} \\ [6pt] 1
\end{array}
\right)
\label{eq:neqm}
\end{equation}
which should be solved for the same number of unknowns $A_k^{(i)}$ at $k=1,2,\ldots,N$ and $\Lambda_i$, where $G_{kk}^{\varepsilon (i)} = G_{kk}^{(i)} + \varepsilon_i \mathcal{R}_i^2$ with
\begin{align}
\label{eq:mela}
G_{kl}^{(i)} &= \frac{1}{M_i} \sum_{j=1}^M { }^{'} \Theta(r_{ij}^\ast-r_{{\rm c},j}) \,
{\bf R}_{ij,k} \bm{\cdot} {\bf R}_{ij,l} \, ,\\
G_k^{(i)} &= \frac{1}{M_i} \sum_{j=1}^M { }^{'} \Theta(r_{ij}^\ast-r_{{\rm c},j}) \,
{\bf R}_{ij,k} \bm{\cdot} {\bf R}_{ij} \, ,
\label{eq:melb}
\end{align}
and $l=1,2,\ldots,N$. The Lagrange multiplier $\Lambda$ normalizes the linear equations according to the constraint $\sum_k A_k = 1$ [see Eq.~(19)]. The latter is necessary to make the extrapolation to be automatically exact for the spatially homogeneous part of the interactions in the transformed space, see the first term in the rhs of Eq.~(17)]. The second term is reproduced approximately by means of coordinate extrapolation (18). Note that the $(N+1) \times (N+1)$ square matrix in Eq.~(25) remains symmetrical since the $\varepsilon_i \mathcal{R}_i^2$-addition concerns only diagonal elements.

The balance contributions $\varepsilon_i \mathcal{R}_i^2 >0$ appear in $G_{kk}^{\varepsilon (i)}$ as a result of the minimization $\sum_k A_k^2=\min$ for the norm of the expansion coefficients [see Eq.~(19)]. Such an additional minimization is needed for the following reason. The dual (virtual coordinate and actual force) extrapolations (with the same expansion coefficients) tentatively assumes that lowering of the coordinate residuals should immediately lead to a decrease of the deviations between the approximated and original forces. But this can be not so in general. For example, when the number $N$ of knots approaches the effective number of local internal degrees of freedom $3M_i$ of neighbouring atoms, the least-square coordinate deviations (the first term in the lhs of Eq.~(20)) will tend to the global (zeroth) minimum. Then some or all coefficients $A_k$ may accept large negative and positive values, despite the imposed normalization $\sum_k A_k=1$. It is well known from the general theory of extrapolation and quadrature formulas that the existence of weights large in magnitude decreases the region of stability, leading to an appreciable increase of the uncertainties outside of it. Moreover, at $N > 3M_i$ the zeroth minimum can be achieved by different sets of $A_k$. The additional minimization $\varepsilon_i \mathcal{R}_i^2 \sum_k A_k^2=\min$ is introduced in Eq.~(20) just to avoid the above singularity and deal with unique solutions at any $N$. Non-zero values of $\varepsilon_i$ allow one to effectively balance between the two kinds of the minimization. Of course, the balancing parameters $\varepsilon_i$ cannot be chosen too large because then the minimization of the squared norm of $A_k$ will be carried more aggressively than that of the coordinate residuals. These parameters should be treated as a small quantity aiming at improving the quality of solvation force extrapolation. Optimal values of $\varepsilon_i$ can be found from actual simulations to obtain the best accuracy.

Another important issue is to provide the coincidence of the extrapolated force $\bm{\mathfrak{\tilde f}}_i(t)$ with one of its knot value $\bm{\mathfrak{f}}_{i,k}$ in situations when the current set of coordinates coincides with one ($k^\ast$-th say) of those related to the reference list, i.e., when ${\bf r}_{ij}(t) = {\bf r}_{ij,k^\ast}$ for all $j \ne i$ at a given $i$. Eq.~(24) says that this is possible provided $A_{k^\ast}^{(i)}=1$ with $A_k^{(i)}=0$ for $k \ne k^\ast$ at each $i=1,2,\ldots,M$. On the other hand, in view of Eq.~(25) the latter conditions can be satisfied if and only if $\mathcal{R}_i = 0$ when $\{{\bf r}_{ij}\}(t) = \{{\bf r}_{ij,k^\ast}\}$. Thus the balance function $\mathcal{R}_i$ should present a measure of minimal deviations of current local configurations $\{{\bf r}_{ij}\}(t)$ from the reference ones $\{{\bf r}_{ij,k}\}$, or ${\bf R}_{ij}$ from ${\bf R}_{ij,k}$ in the transformed space. Replacing ${\bf R}_{ij}$ by ${\bf R}_{ij}^\ast \equiv \bm{\varrho}_{ij}^\ast$ according to the frequency reuse regime, the obvious expression for this measure is
\begin{equation}
\mathcal{R}_i^2(t) = \min_{k=1}^N \left[
\frac{1}{M_i} \sum_{j=1}^M { }^{'} \Theta(r_{ij}^\ast-r_{{\rm c},j})  \left( {\bf R}_{ij,k} -
{\bf R}_{ij}^\ast \right)^2 \right] \equiv \min_{k=1}^N \mathcal{R}_{ik}^2 \equiv \min_{k=1}^N \lambda^\textrm{(1)}_{i,k} \, .
\end{equation}
Here the expression under minimization coincides with the lhs of Eq.~(11), so the balancing function can be represented as the smallest eigenvalue $\lambda_1 \equiv \lambda^\textrm{(1)}_{i,k}$ among all $k=1,2,\ldots,N$. In such a way we come to the so-called dynamical balancing scheme where the force extrapolation provides exact results in the limits when the local current spatial configuration is very close (or coincides, then $\mathcal{R}_i = 0$) with one of those containing in the reference set. This is in a contrast to previous approximation schemes \cite{Omelyan:2013:139:244106,Omelyann} which apply the static balancing ($\mathcal{R}_i \equiv {\rm const}$) and, thus, distort the true values of solvation forces in such limits.

\textit{Extending the reference list.} --- The accuracy of the force extrapolation will improve with increasing the number of points $N$ in the reference list. However, we cannot put $N$ to be too large because then the number of linear equations increases, too. These equations (25) need to be solved frequently even in the frequency reuse regime, namely, $N M p$ times, where $1 \ll p \ll P$ with $P=h/\Delta t \gg 1$, reducing the efficiency of the extrapolation. A way to avoid this lies in the following. We can extend the reference list from a relatively small number of $N \lesssim 100$, say, to a larger value $N' \gg N$ by collecting the reference coordinates ${\bf r}_{ij,k'}$ and forces $\bm{\mathfrak{f}}_{i,k'}$ with $k'=1,2,\ldots,N'$ during a wide previous time interval $\Delta H = N' h \gg N h$. Then the squared distances $\mathcal{R}_{ik'}^2 \equiv \lambda^\textrm{(1)}_{i,k'}$ in the $3M$-dimensional space between the transformed outer coordinates $\{{\bf R}_{ij,k'}\}$ and the current origin point $\{{\bf R}_{ij}^\ast(t)\}$ (see Eq.~(28)) can be sorted in the ascending order with respect to $k'$ at a given $i$, and the first $N$ most closest (to ${\bf R}_{ij}^\ast$) points can be selected among the extended set to satisfy the condition $\mathcal{R}_{i1} < \mathcal{R}_{i2} < \ldots < \mathcal{R}_{i N}$. The forces $\bm{\mathfrak{f}}_{i,k'}$ must be resorted synchronically with the coordinates ${\bf r}_{ij,k'}$ to form the basic reference list, i.e., the best pair subset with $N$ points. It should then be used when performing the force extrapolation (24).

The above procedure can further improve the quality of the extrapolation, especially at $N' \gg N$. The reason is that the choice of the nearest outer pairs in the transformed space additionally reduces the coordinate region in which the extrapolation is performed. This leads to a decrease of the coordinate residuals and, as a consequence, to an increase of the accuracy. In fact, such an additional reduction minimizes the change in the transformed solvation forces during torsion motions of the solute macromolecule. Note that such motion (characterizing by large amplitudes) is responsible for transitions of the biomolecule from one conformational pool to another where the torsion potential has a local minima. Thus, an optimal value for the expanded interval $\Delta H = N' h$ should be of order of the mean life time in local conformational minima. Then, whenever the transition to other conformations occurs, we can quickly reselect the subset to fit the basic outer points to the current solute conformation. The accuracy of such fitting is especially high if the molecule has already been near this conformation at previous times.

Worth remarking is that the selecting procedure at $N \ll N'$ requires only little extra numerical efforts even for large enough $N' \gtrsim 10^3$. The reason is that the computational cost grows linearly with $N'$ (at a tiny proportionality factor). Indeed, the selection needs to know only the lowest eigenvalues $\lambda^\textrm{(1)}_{i,k'}$ of small $4 \times 4$ matrices and not their eigenvectors. The latter ${\bf q}^\textrm{(1)}_{i,k}$ are necessary only for the best subset with $k=1,2,\ldots,N \ll N'$ to build the rotation matrix ${\bf S}_{i,k}$ for the force extrapolation (24). On the other hand, the computational efforts increase much more rapidly with $N$, namely, proportionally to $(N+1)^3$, in order to find solutions to $(N+1)$ linear equations (25). Note also that the selection procedure is performed only once per many ($p \gg 1$) inner time steps, further lowering the numerical expenses.

\textit{Resulting algorithm.} --- In view of the techniques introduced in the preceding subsections, the resulting enhanced solvation force extrapolation algorithm can be briefly described as follows. At the very beginning, the 3D-RISM-KH integral equations are solved after each $\Delta t$ of the $N$ first inner steps with no extrapolation to fill out the basic reference list. Then the extrapolation starts with $N$ points and the extended list is accordingly completed step by step during the time integration to achieve the maximal length with $N'>N$ pairs. Since $h$ can be much larger than $\Delta t$, we cannot put the outer step to be immediately equal to $h \gg \Delta t$. The reason is that this skews the extrapolation because of the significant non-uniformity of the time intervals between the points from the list. That is why the outer time interval should be increased smoothly every inner step from $\Delta t$ until $h$ with an increment of $\Delta t$.

Further, after each $p$ inner steps, we solve the eigenvalue problem (16) for the extended list with $N'$ coordinates. The first best $N<N'$ points are selected by sorting the corresponding smallest eigenvalues in the ascending order. The coordinates and forces related to the best subset obtained are then transformed by individual non-Eckart rotation transformations in terms of the ${\bf S}$-matrix (13) constructed on the $N$ smallest eigenvectors. Having the transformed coordinates, we build the system of $(N+1)$ linear equations (25) and solve it for the expansion coefficients by an inversion of the $(N+1) \times (N+1)$ matrix (26). The inversion is carried out only once during the time interval $p \Delta t$ because then this matrix remains unchanged, while the right-hand side vector in (25) varies every $\Delta t$ according to Eq.~(27).

Using the expansion coefficients, the solvation forces are extrapolated at each inner step $\Delta t$ within the outer time interval $t \in ]t_1, t_0[$ of length $h$ as the weighted sum of their $N$ previous outer transformed values, followed by the inverse transformation (24). The extrapolation procedure is applied $h/\Delta t$ times to achieve the next outer point. At that point, the solvation forces are again calculated explicitly by solving the 3D-RISM-KH integral equations. The extended $N'$-list is then updated by the new outer force-coordinate pair, while the oldest one is discarded. All these actions are repeated $H/h$ times for the next outer intervals until the desired simulation time length $H$ is achieved.

This completes the derivation of the enhanced solvation force extrapolation (ESFE) algorithm.

\section{Solving the ESFE/3D-RISM-KH equations of motion}

\textit{Combining ESFE/3D-RISM-KH with MD.} --- The first issue in combining the ESFE/3D-RISM-KH approach with the method of MD is the choice of statistical ensemble in which the dynamics of the system will be considered. As was shown in previous MD simulations at the presence of extrapolated 3D-RISM-KH solvation forces, the best ensemble in context of stability and efficiency allowing large inner and outer time steps is the so-called optimized isokinetic Nos\'e-Hoover chain (OIN) thermostat. \cite{Omelyan:2013:39:25, Omelyan:2013:139:244106, Omelyann} The equations of motion for solute atoms in hybrid MD/ESFE/3D-RISM-KH simulations
in the OIN ensemble steered with 3D-RISM-KH mean solvation forces which are extrapolated with the ESFE approach can be cast in the following compact form
\begin{equation}
\frac{d \bm{\Gamma}}{d t} = L \bm{\Gamma}(t) \, ,
\label{eq:lem}
\end{equation}
where $\bm{\Gamma}=\{ {\bf r}, {\bf v}; \bm{\sigma}, \bm{\nu} \}$ denotes the extended phase space and $L$ is the Liouville operator. The extended space, apart from the full set of coordinates ${\bf r} \equiv \{ {\bf r}_i \}$ and velocities ${\bf v} \equiv \{ {\bf v}_i \}$ of all solute atoms, includes also all thermostat frequencies $\bm{\nu} \equiv \{ \nu_{\kappa,i} \}$ with $\kappa=1,\ldots,\mathcal{K}$ and their conjugated dynamical variables $\bm{\sigma} \equiv \{ \sigma_i \}$. The latter are introduced by means of the relation $d \sigma_i/dt= (\tau_i^2 \nu_{1,i}^2 \nu_{2,i}-\sum_{\kappa=2}^\mathcal{K} \nu_{\kappa,i})$, where $\mathcal{K}$ is the number of chains per thermostat. The Liouvillian can be split up as $L = \sum_{i=1}^M ({\mathcal A}_i + {\mathcal B}_i + {\mathcal C}_{v,\nu,i} + {\mathcal C}_{\nu,i} + {\mathcal C}_{\sigma,i})$ into the kinetic ${\mathcal A}_i = {\bf v}_i \bm{\cdot} \partial/\partial {\bf r}_i$, potential
\begin{equation}
{\mathcal B}_i = \bigg( \frac{{\bf f}_i}{m_i} - {\bf v}_i
\frac{{\bf v}_i \bm{\cdot} {\bf f}_i}{2 {\rm T}_i} \bigg)
\bm{\cdot} \frac{\partial}{\partial {\bf v}_i} -
\frac{ {\bf v}_i \bm{\cdot} {\bf f}_i}{2 {\rm T}_i}
\nu_{1,i} \frac{\partial}{\partial \nu_{1,i}} \, ,
\label{eq:lpt}
\end{equation}
and chain-thermostat parts with
\begin{equation}
{\mathcal C}_{v,\nu,i} = \frac{\tau_i^2 \nu_{1,i}^2}{4} \nu_{2,i}
{\bf v}_i \bm{\cdot} \frac{\partial}{\partial {\bf v}_i} +
\bigg( \frac{\tau_i^2 \nu_{1,i}^2}{4} - 1 \bigg)
\nu_{1,i} \nu_{2,i} \frac{\partial}{\partial \nu_{1,i}} \, ,
\label{eq:lcvw}
\end{equation}
\begin{equation}
{\mathcal C}_{\nu,i} = \sum_{\kappa=2}^\mathcal{K} \Big(
\nu_{\kappa-1,i}^2-\frac{1}{\tau_i^2}-\nu_{\kappa+1,i}
\nu_{\kappa,i} \Big) \frac{\partial}{\partial \nu_{\kappa,i}} \, ,
\label{eq:lcw}
\end{equation}
\begin{equation}
{\mathcal C}_{\sigma,i} = -\bigg( \tau_i^2 \nu_{1,i}^2 \nu_{2,i} -
\sum_{\kappa=2}^\mathcal{K} \nu_{\kappa,i} \bigg)
\frac{\partial}{\partial \sigma_i} \, .
\label{eq:lcv}
\end{equation}
Mention that in the canonical OIN ensemble \cite{Omelyan:2013:39:25} each atom is coupled with its own thermostat by imposing the constraint ${\rm T}_i=3 k_{\rm B} T/2$, where ${\rm T}_i=m_i {\bf v}_i^2/2 + 3k_{\rm B} T/4 \tau_i^2 \nu_{1,i}^2/2$ is the full kinetic energy of the $i$-th subsystem. The quantity $\tau_i$ is related to the relaxation time, determining the strength of coupling of atom $i$ with its thermostat.

The total forces ${\bf f}_i = \bm{f}_i + \bm{\mathfrak{f}}_i$ are now divided into the fast (f) solute-solute component $\bm{f}_i$ and slow (s) 3D-RISM-KH solute-solvent one $\bm{\mathfrak{f}}_i$. In view of Eq.~(30), this results in the corresponding splitting of the potential operator as ${\mathcal B}_i(\{{\bf f}_i\}) = {\mathcal B}_i(\{\bm{f}_i\}) + {\mathcal B}_i(\{\bm{\mathfrak{f}}_i\}) \equiv {\mathcal B}_{\rm f} + {\mathcal B}_{\rm s}$. Remember that the instantaneous solute-solute forces $\bm{f}_i$ are calculated always directly (by $-\partial \mathcal{U}/\partial {\bf r}_i$), while the 3D-RISM-KH mean solvation forces $\bm{\mathfrak{f}}_i$ are either evaluated explicitly [see Eq.~(7)] or approximated by $\bm{\mathfrak{\tilde f}}_i$ using ESFE (24). In the latter case ${\mathcal B}_{\rm s}(\{\bm{\mathfrak{f}}_i\})$ transforms to ${\mathcal B}_{\rm s}(\{\bm{\mathfrak{\tilde f}}_i\}) \equiv \tilde {\mathcal B}_{\rm s}$.

\textit{Multiple time step decompositions in OIN.} --- Acting in the spirit of the multiple time step (MTS) decomposition method, \cite{Omelyan:2011:135:114110, Omelyan:2011:135:234107, Omelyan:2012:8:6, Omelyan:2012:85:026706} the solution $\bm{\Gamma}(h)=e^{L h} \bm{\Gamma}(0)$ to Eq.~(29) over the outer time interval $h$ from an initial state $\bm{\Gamma}(0)$ can be presented \cite{Omelyan:2013:39:25} as the following product of exponential operators:
\begin{equation}
\bm{\Gamma}(h)=\prod_{n'=1}^n e^{{\mathcal C} \frac{\delta t}{2}}
e^{{\mathcal B}_{\rm fs}^{(n')} \frac{\delta t}{2}} e^{{\mathcal A} \delta t}
e^{{\mathcal B}_{\rm fs}^{(n')} \frac{\delta t}{2}} e^{{\mathcal C} \frac{\delta t}{2}}
\bm{\Gamma}(0) + \mathcal{O}(\delta t^2) \, .
\label{eq:dec}
\end{equation}
Here, $n=h/\delta t \gg 1$ is the total number of sub-inner time steps with length $\delta t \ll \Delta t$ each, ${\mathcal C} = {\mathcal C}_{v,\nu} + {\mathcal C}_\nu + {\mathcal C}_\sigma$,
\begin{equation}
e^{{\mathcal B}_{\rm fs}^{(n')} \frac{\delta t}{2}} = \left\{
\begin{array}{ll}
e^{{\mathcal B}_{\rm f} \frac{\delta t}{2}} e^{{\mathcal B}_{\rm s} \frac{\Delta t}{2}} \, ,
\ &\ \ \ \text{only once per $h$ when $n'/\frac{\Delta t}{\delta t}=1$}
\\ [4pt]
e^{{\mathcal B}_{\rm f} \frac{\delta t}{2}} e^{\tilde {\mathcal B}_{\rm s} \frac{\Delta t}{2}} \, ,
\ & \ \ \ \text{for other inner steps, $n'=2\frac{\Delta t}{\delta t},\ldots,n$}
\\ [4pt]
e^{{\mathcal B}_{\rm f} \frac{\delta t}{2}} \, , \ & \ \ \ \text{for all rest $n'$}
\end{array}
\right.
\label{eq:gvo}
\end{equation}
is the generalized velocity propagator, $\Delta t$ is the inner ($\delta t \ll \Delta t \ll h$) time step, $\mathcal{O}(\delta t^2)$ is the accuracy of the decomposition, and the subscript $i$ is omitted for the sake of simplicity. Note that we should first update (by $e^{{\mathcal C} \delta t/2}$ and $e^{{\mathcal B} \delta t/2}$) the complete set of velocities ${\bf v}_i$ and frequencies $\nu_{\kappa,i}$ belonging to all atoms ($i=1,2,\ldots,M$) and thermostat chains ($\kappa=1,\ldots,\mathcal{K}$) before to change the coordinates ${\bf r}_i$ of all particles by $e^{{\mathcal A} \delta t}$. A nice feature of the OIN decomposition is that the action of all the single-exponential operators which arise in Eqs.~(34) and (35) on $\bm{\Gamma}$ can be handled analytically using elementary functions. \cite{Omelyan:2013:39:25}

Therefore, the propagation $\bm{\Gamma}(t)=[\bm{\Gamma}(h)]^{t/h}$ of dynamical variables from their initial values $\bm{\Gamma}(0)$ to arbitrary time $t=H>0$ in future can be performed by consecutively applying the single exponential transformations of a phase space point $\bm{\Gamma}$ in the order defined in Eq.~(34). As can be seen in Eq.~(35), the fastest ${\mathcal B}_{\rm f}$-component of motion is integrated most frequently, namely, $n=h/\delta t$ times per outer interval $h$ with the smallest (sub-inner) time step $\delta t$, while the (original or approximated) slow 3D-RISM-KH forces are applied impulsively only every $\Delta t/\delta t$ sub-inner step, i.e., $h/\Delta t<n$ times. Note that almost all these impulses (when $\Delta t \ll h$) are obtained by employing the extrapolated 3D-RISM-KH forces [Eq.~(24)] in terms of operator $\tilde {\mathcal B}_{\rm s}$, while the explicit 3D-RISM-KH calculations (7) are used in ${\mathcal B}_{\rm s}$ only once per outer time interval $h$. Taking into account that the solute-solute forces are much cheaper to evaluate than the solvation ones, obvious speedup is achieved as compared to the single time-stepping propagation ($n=1$, $\delta t= \Delta t$) without extrapolation ($\Delta t=h$). Furthermore, the existence of the impulsive inner time steps of length $\Delta t > \delta t$ gives a possibility of reducing the number of (either extrapolative or direct) 3D-RISM-KH evaluations from $h/\delta t$ to $h/\Delta t$. Finally, applying the ESFE approach allows further significant improvement of the overall efficiency, since the most expensive 3D-RISM-KH calculations are performed just once per outer step $h$.

In view of the above, the following hierarchy of time steps
\begin{equation}
\delta t \ll \Delta t \ll h \ll N h \ll N' h = \Delta H \ll H
\end{equation}
should be set in order to achieve an optimal performance of hybrid MTS-MD/OIN/3D-RISM-KH simulations using the ESFE approach. This completes coupling of ESFE with MD. We will refer to the resulting scheme as a hybrid MTS-MD/OIN/ESFE/3D-RISM-KH method, or simply OIN/ESFE/3D-RISM for brevity.

Mention that the quasidynamic obtained in MTS-MD/OIN/ESFE/3D-RISM-KH simulations will differ from the true dynamics of conventional MD with explicit solvent. In particular, such quasidynamics does not obey the Maxwell velocity distribution and, thus, unlike microcanonical MD, cannot get us real time correlation functions. However, as was rigorously proven, \cite{Omelyan:2013:39:25} the configurational part of the extended partition function related to MTS-MD/OIN/3D-RISM-KH simulations at targeted temperature $T$ does coincide with the true canonical distribution of the physical system in coordinate space. This is a very important feature because the original conformational properties, including spatial atom-atom density distribution functions, can then be readily reproduced. Such quasidynamical sampling appears to be much more efficient than that following from ``real-time'' (microcanonical or canonical) brute-force MD simulations (because of excluding slow solute-solvent re-equilibration precesses, see Section 2).

\begin{figure}
\centering
\includegraphics[width=0.74\textwidth]{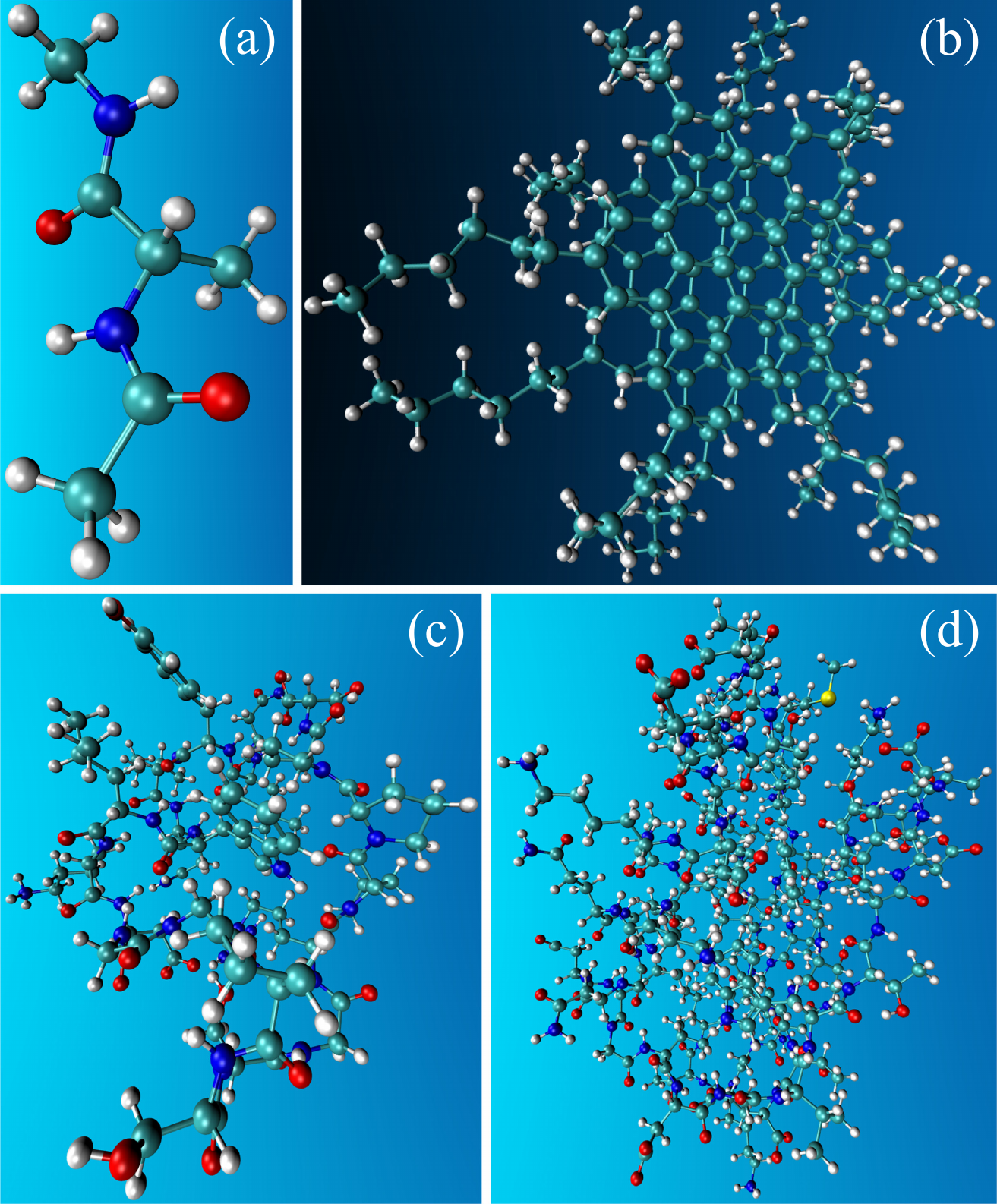}
\caption{Schematic ball-and-stick representation of the molecular structures corresponding to (a) alanine dipeptide ($M=22$ atoms), (b) asphaltene ($M=336$), (c) miniprotein 1L2Y ($M=304$), and (d) protein G ($M=862$). Different types of atoms are shown by the following colors: white--Hydrogen, red--Oxygen, green--Carbon, blue--Nitrogen and yellow--Sulfur.}
\label{fig:MolStruct}
\end{figure}

\section{Application of MTS-MD/OIN/ESFE/3D-RISM-KH simulations to solute-solvent liquids}

\textit{Numerical details.} --- The proposed MTS-MD/OIN/ESFE/3D-RISM-KH approach will now be validated in actual simulations. First of all, the following six source files: \verb"SANDER.F", \verb"RUNMD.F", \verb"MDREAD.F", \verb"MD.H", \verb"AMBER_RISM_INTERFACE.F" and \verb"FCE_C.F" were taken from the original Amber 2018 package \cite{Amber} and accordingly altered to implement the approach into the parallel program code. The systems considered are: (i) fully flexible model of hydrated alanine dipeptide ($M=22$ atoms), (ii) asphaltene ($M=336$) in toluene (C$_6$H$_5$CH$_3$), as well as (iii) 1L2Y-miniprotein ($M=304$), and (iv) protein G ($M=862$) both solvated in water (H$_2$O). Schematic representation of the molecular structures related to these four solute macromolecules are presented in Fig.~1. The Amber03, \cite{Duan:2003:24:1999} Amber99SB \cite{Simmerling:2002:124:11258} and general Amber \cite{Wang:2004:25:1157} force fields were used to model the interactions in alanine dipeptide and miniprotein 1L2Y, in protein G and asphaltene, respectively. Water was described by the modified cSPC/E model. \cite{Kovalenko:2003:169, Miyata:2008:29:871, Luchko:2010:6:607} The interaction constants for toluene solvent were extracted from optimized potentials \cite{Jorgensen:1993:14:206} of the general Amber force field. We applied free boundary conditions and an adaptive solvation box with varying sizes determined by the current diameter of the solute molecule plus a buffer space of width $R_{\rm b}=10$~\AA. Note that the mean diameters of the alanine dipeptide, asphaltene, miniprotein 1L2Y, and protein G macromolecules are about 9, 28, 26, and 42~\AA, respectively. The cutoff radius of the solute-solvent interactions was set to $R_{\rm c}=14$~\AA. No truncation was made for the solute-solute forces. The 3D-RISM-KH integral equations were discretized on a rectangular grid with resolution $\delta r=0.5$~\AA\ and converged to a relative root mean square residual tolerance of $\delta \epsilon = 10^{-4}$ using the MDIIS algorithm. \cite{Kovalenko:2003:169} Further increase of $R_{\rm c}$ and $R_{\rm b}$, as well as decrease of $\delta r$ and $\delta \epsilon$ did not noticeably affect the results.

\begin{table}
\caption{Sets of some parameters used in MTS-MD/OIN/ESFE/3D-RISM-KH simulations of different systems.}
\def\arraystretch{1.4}
\begin{center}
\begin{tabular}{|c|c|c|c|c|c|c|c|c|c|c|c|}
\hline
\ \ \ \ \ System \ \ \ \ \ & \ \ \ $M$ \ \ \ & \ \ \ \ \ \ \ \ $\eta$ \ \ \ \ \ \ \ \ & \ \ \ weight \ \ \ & \ \ \ \ $r_{\rm c}$ \ \ \ \ & \ \ \ \ $\varepsilon$ \ \ \ \ & \ \ \ \ $p$ \ \ \ \ & \ \ \ \ $N$ \ \ \ \ & \ \ \ $N'$ \ \ \ & \ \ \ \ $\tau$ \ \ \ \ & \ \ \ $\mathcal{K}$ \ \ \ & \ \ \ \ $h_{\rm m}$ \ \ \ \ \\
\hline \hline
alanine & $22$ & $0.7$ \AA$^{-1}$ & charge & \ $6$ \AA \ & $0.1$ & \ $5$ \ & \ $56$ \ & $4000$ & $10$ fs & $2$ & $32$ ps \\
\hline
asphaltene & $336$ & $0.85$ \AA$^{-1}$ & \ mass \ & $14$ \AA & $0.5$ & $25$ & \ $36$ \ & $4000$ & $20$ fs & $4$ & $8$ ps \\
\hline
miniprotein & $304$ & \ $0.7$ \AA$^{-1}$ \ & \ force \ & $14$ \AA & $0.1$ & $25$ & \ $56$ \ & $1000$ & $40$ fs & $8$ &  \ $8$ ps \\
\hline
protein G & $862$ & \ $0.7$ \AA$^{-1}$ \ & \ force \ & $14$ \AA & $0.1$ & $25$ & \ $56$ \ & $1000$ & $40$ fs & $8$ & \ $4$ ps \\
\hline
\end{tabular}
\end{center}
\end{table}

The number $M$ of atoms per macromolecule, optimal values for the exponential scaling parameter $\eta_{ij} \equiv \eta$, cut-off radius $r_{\rm c}$ and balance parameter $\varepsilon_i \equiv \varepsilon$, as well as the frequency number $p$, main basic and extended reference list lengths $N$ and $N'$ used in MTS-MD/OIN/ESFE/3D-RISM-KH simulations for each system are presented in Table~1. Optimal types of the weighting scaling scheme, the numbers $\mathcal{K}$ of chains, relaxation times $\tau \equiv \tau_i$ of the OIN thermostat and maximal allowed outer time steps $h_{\rm m}$ are also given there. The sub-inner and inner time steps in the MTS integration were always equal to $\delta t=1$~fs and $\Delta t=8$~fs, respectively. Up eleven MD series with different values of the outer time step, namely, $h=12$, 24, 96, 200, 400~fs, 1, 2, 4, 8, 16, and 32~ps have been carried out within the OIN/ESFE/3D-RISM-KH approach to obtain a whole pattern (see figures below) on accuracy of the enhanced solvation force extrapolation. In each this series, the numbers of points of the basic and extended sets vary in the ranges $1 \le N \le 100$ and $N \le N' \le 4000$. The total duration of the simulations was $H=25-160$~ns in dependence on the system. For the purpose of comparison with ESFE, the results of the SFE, \cite{Luchko:2010:6:607} ASFE \cite{Omelyan:2013:139:244106} and GSFE \cite{Omelyann} extrapolation schemes have been prepared as well.

The runs were performed at a temperature of $T=300$~K and a solvent density of $\rho = 1$~g/cm$^3$ for water and $0.87$~g/cm$^3$ for toluene. The simulations of miniprotein and protein G started from the folded crystal conformations obtained in NMR experiment, taken from PDB (protein data bank) structures 1L2Y \cite{Neidigh:2002:9:425} and 1P7E, \cite{Ulmer:2003:125:9179} respectively. The initial structure of the asphaltene dimer was based on the full geometry optimization using density functional theory at the $\omega$B97X-D/6-31G* level. \cite{Chai:2008:10:6615} The conventional canonical MD simulations of hydrated alanine dipeptide and miniprotein in explicit solvent were carried out, too, using Amber 2018 and involving the SPC/E \cite{Berendsen:1987:91:6269} and TIP3P \cite{Jorgensen:1983:79:926} models of water with 1263 and 16895 molecules at $R_{\rm c}=14$~\AA\ and $8$~\AA\ cutoffs in the direct space for nonbonded electrostatic interactions. The truncation terms were handled by the particle-mesh Ewald summation method \cite{Essmann:1995:107:113} with periodic boundary conditions. The equations of motion were solved with a single time step of $\delta t = \Delta t=2$~fs (and no extrapolation) exploiting the Langevin dynamics \cite{Loncharich:1992:32:523} at a friction viscosity of $\gamma=1$~ps$^{-1}$ as well as SHAKE \cite{Ryckaert:1977:23:327, Ciccotti:1982:47:1253} to fix hydrogen bonds.

\textit{Results for extrapolation accuracy.} --- The accuracy of the extrapolation was estimated by measuring the relative mean square deviations
\begin{equation}
\Psi=\frac12 \frac{\left \langle \sum_{i=1}^M (\bm{\mathfrak{\tilde f}}_i - \bm{\mathfrak{f}}_i)^2 \right \rangle^{1/2}}{\left \langle \sum_{i=1}^M \bm{\mathfrak{f}}_i^2
\right \rangle^{1/2}}
\label{est}
\end{equation}
of the approximated [Eq.~(24)] forces ${\bf \tilde f}_i$ from their original values ${\bf f}_i$ [calculated explicitly via the 3D-RISM-KH relation (7)] at each outer time step, where $\langle \ \ \rangle$ denotes the statistical averaging along the whole simulation length. Note that during each $h$, the the deviations increase from zero at the very beginning (when the inner coordinates coincide with those of the first basic point) to maximal values at the end of the current outer time interval, so that the fraction $1/2$ is necessary to get mean values. It is worth remarking also that such an estimation does not require any extra computational efforts, since it operates with outer forces which are already known during evaluation of the equations of motion.

\begin{figure}
\centerline{\includegraphics[width=0.85\textwidth]{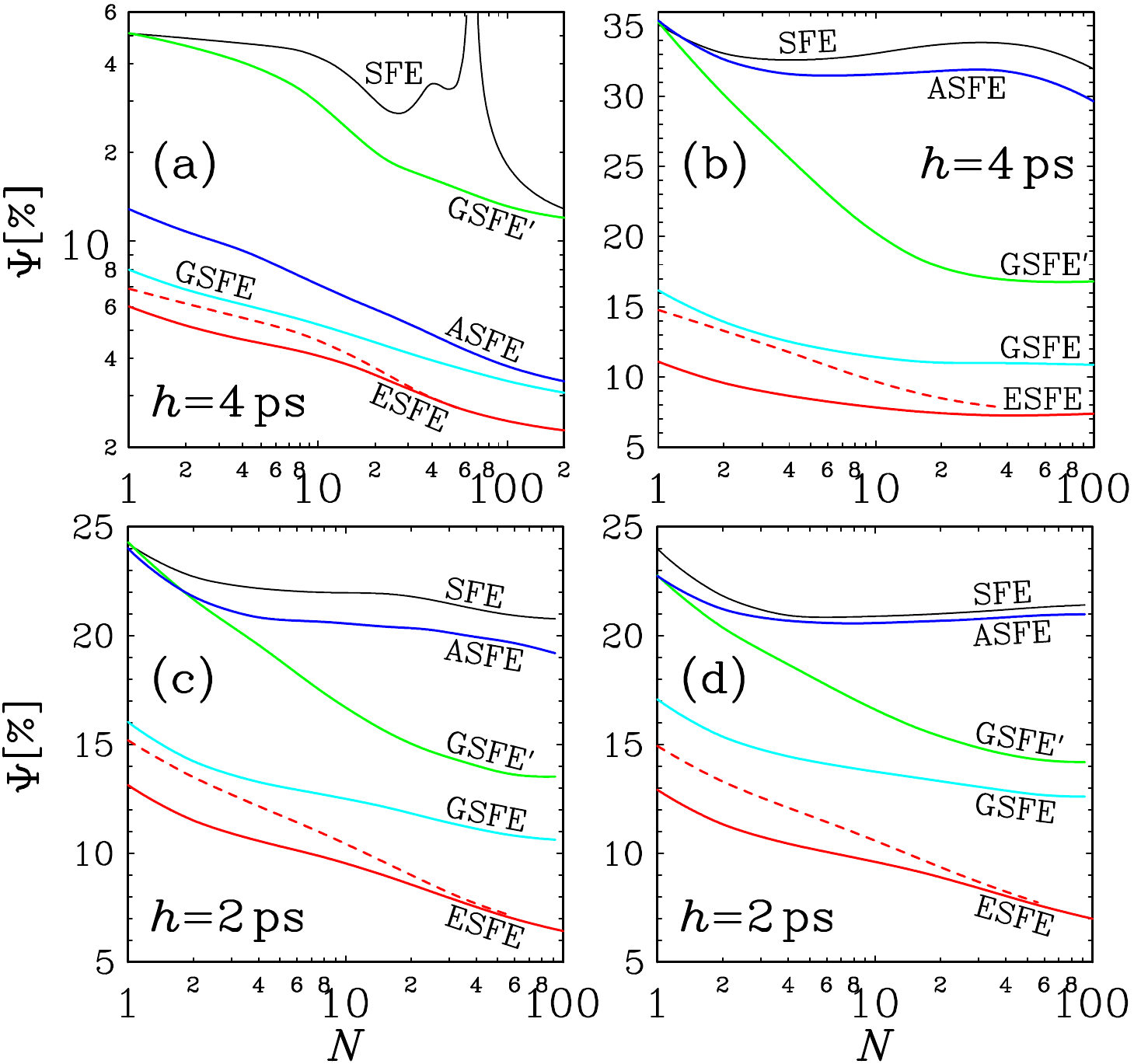}}
\caption{Uncertainties $\Psi$ of the solvation force approximation as functions of the number$N$ of basic points in the reference list corresponding to the MD/OIN/3D-RISM-KH simulations of hydrated alanine dipeptide (a), asphaltene in toluene (b), as well as miniprotein (c) and protein G (d) both in water using different extrapolation approaches [see the text] at outer time steps of $h=4$ or 4~2s.}
\end{figure}

The relative mean square deviations $\Psi$ obtained in the MD/OIN/3D-RISM-KH simulations of hydrated alanine dipeptide, asphaltene in toluene, hydrated miniprotein, and protein-G in water using various extrapolation approaches at most characteristic outer time steps $h=2$ and $4$~ps are shown in parts (a), (b), (c), and (d) of Fig.~2, respectively, versus the number $N$ of the basic points (at a given $N'=4000$ or $1000$, see Table~1). These approaches are: the standard solvation force extrapolation (SFE) scheme, \cite{Luchko:2010:6:607} advanced SFE (ASFE) of Ref.~\onlinecite{Omelyan:2013:139:244106}, generalized SFE (GSFE) of Ref.~\onlinecite{Omelyann}, as well as the enhanced SFE (ESFE) proposed in the present paper. The SFE, ASFE, GSFE, and ESFE functions $\Psi(N)$ are plotted by the black, blue, cyan, and red curves, respectively. The latter are solid (regular regime, $p=1$) or dashed (frequency reuse regime, $p \gg 1$, Table~1). A variant GSFE$'$ of GSFE with global (instead of individual) rotational transformations is also included (green curves). Remember that SFE uses only the least-square minimization (in the basic reference list) with no extension, normalization, weighting, balancing, transformation and truncation. In ASFE, the global rotational transformations, static balancing and extension of the reference list are included additionally. GSFE considers in addition the individual (instead global) rotations as well as simple weighting and truncation. Finally, the ESFE method additionally contains the exponential scaling transformation with different weighting schemes and the dynamical (instead static) balancing during the minimization.

As can be seen, the SFE approach leads to the worst accuracy of the force extrapolation with the largest deviations $\Psi$ for any values of $N$ and $h$. Moreover the SFE function $\Psi(N)$ exhibits a singularity (described in Section~3) at $N \sim 3M$ which can be observed in Fig.~2a for alanine dipeptide, where $M=22$. Analogous SFE-singularity exist for other macromolecules, where $M \ge 304$ (they are not presented in Fig.~2a--d merely because $N \le 100$ there). The ASFE method improves the SFE results only for alanine dipeptide and removes the singularity (by including the balancing), while for asphaltene and proteins they remain practically the same. The reason is that ASFE applies global rotation transformations with involving all $M$ atoms of the solute molecule without truncation. The global rotations (caused by the interactions with solvent and thermostat) as a whole are significant only for small solute macromolecules, like alanine dipeptide. With increasing the number $M$ of atoms,
the amplitude of these diffusion-like rotations decreases, lowering the efficiency of the global transformations. This efficiency can be somewhat improved by including simple weighting and truncation, see the curves marked as GSFE$'$. Further improvement in the extrapolation precision can be reached by applying the individual rotation transformations within GSFE, where values of $\Psi(N)$ are reduced appreciably. The best precision of the solvation force approximation for all the systems is obtained within the ESFE method. Here the deviations $\Psi(N)$ between the exact and extrapolated values accept minimal values at each given $N$ (see Fig.~2). In particular, $\Psi=2\%$, $7.6\%$, $6\%$, and $7\%$ at $N = 96$ for alanine dipeptide, asphaltene, miniprotein, and protein G, respectively. Moreover, the ESFE function $\Psi(N)$ continues to decrease at $N > 100$, while the GSFE approach exhibits a saturation in this range. Therefore, the exponential scaling linearization with expanded weighting schemes and the dynamical balancing technique used in this approach indeed allow to decrease the uncertainties to the lowest possible level. At $N \gtrsim 36$ this level almost does not change with increasing the frequency reuse number $p$ from $1$ up to $25$ (take a look at solid and dashed red curves). This is very important feature since the computational costs are smaller at larger $p$.

\begin{figure}[t]
\centerline{\includegraphics[width=0.95\textwidth]{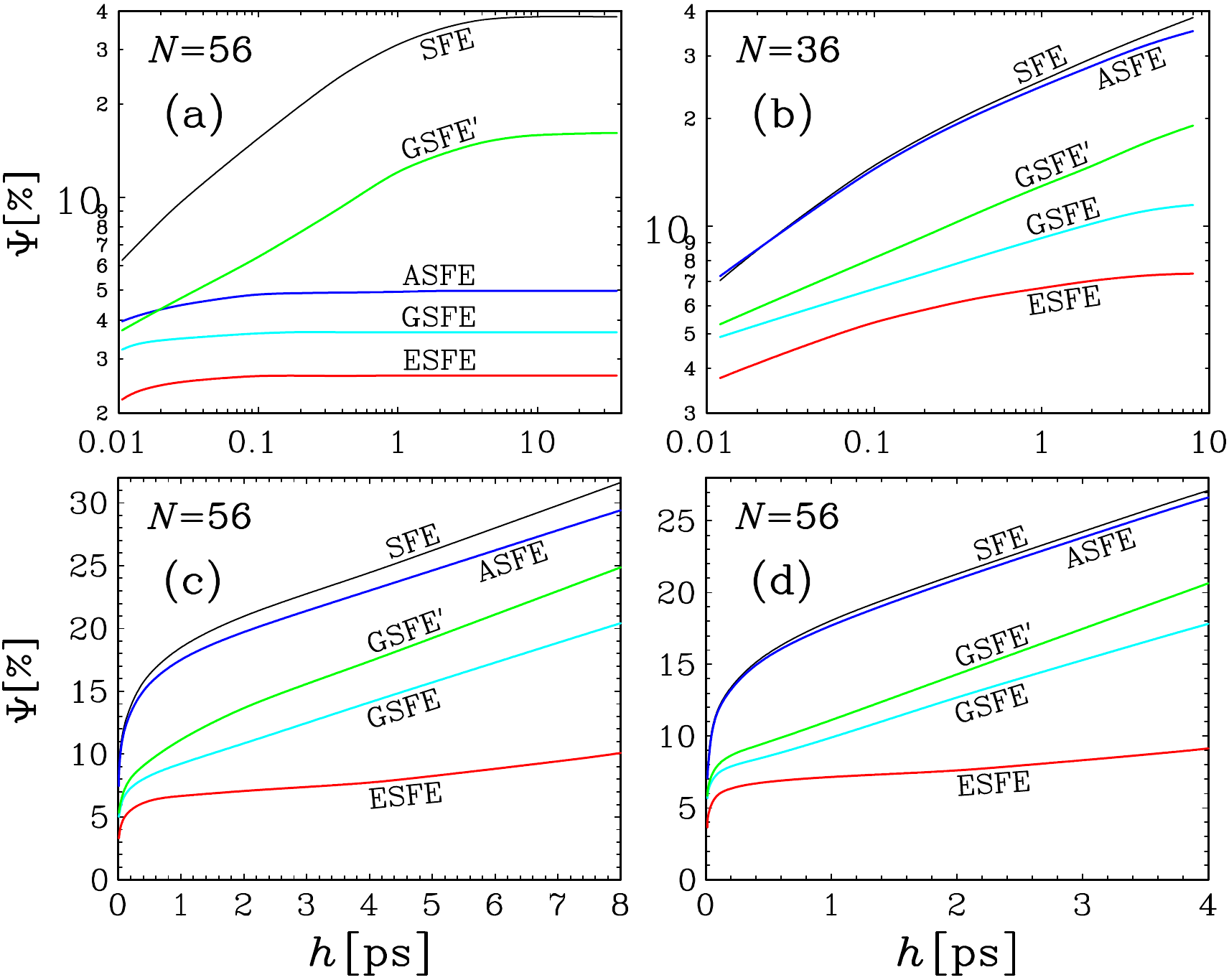}}
\caption{Uncertainties $\Psi$ of the solvation force approximation as depending on the size of the outer time step $h$ regarded to the MD/OIN/3D-RISM-KH simulations of hydrated alanine dipeptide (a), asphaltene in toluene (b), as well as miniprotein (c) and protein G (d) both in water using different extrapolation approaches [see the text] at fixed $N=56$ or $36$.}
\end{figure}

The relative uncertainties $\Psi$ of the solvation force approximation as depending on the size of the outer time step $h$ at given optimal values $N=36$ or $56$ and $N'=4000$ or $1000$ (Table~1) is depicted in Fig.~3 for the four systems and five extrapolation methods considered. Looking at this figure we can say nearly the same words as those presented above when analyzing Fig.~2. Namely, for each system and the same $h$, the differences $\Psi(h)$ decrease when arranging the methods in the following order: SFE, ASFE or GSFE$'$, GSFE, and ESFE, confirming an evident superiority of the latter over all the rest schemes. Huge sizes of the outer time step up to of order of $h \sim 4-32$~ps can be applied within the ESFE extrapolation, maintaining a significantly high accuracy of $\Psi \sim 2 - 10\%$. For example, in the case of the miniprotein, the same ESFE deviation $\Psi \sim 10\%$ at $h=8$~ps is obtained by GSFE at considerably shorter values of $h \sim 1.5$~ps (see Fig.~3c). This means that the MTS-MD/OIN/ESFE/3D-RISM-KH simulations can be accelerated up to $8/1.5 \sim 5$ times with respect to those of GSFE. Similar speedup can be observed for other three systems (Fig.~3a,b,d). It should be pointed out also that the ESFE function $\Psi(h)$ is much more flat than in the case of the SFE, ASFE, and GSFE$'$ approaches. This gives the potential possibility to use ESFE even with longer outer time steps. In particular, for alanine dipeptide the extrapolation deviations $\Psi(h)$ are independent of $h$ at large enough outer time steps (see Fig.~3a). This can be explained by the fact that such a system is characterized by a small number of equilibrium states with relatively short times of life in them, namely, of order of nanoseconds. As a result, the extended reference list with large $N'=4000$ contains almost all important conformations already at $h \ge 1$~ps for ESFE because then $\Delta H = N' h \sim 4$~ns. With increasing the complexity of the macromolecule, the number of equilibrium states and the lifetimes in them grows rapidly. At finite $N' \sim 1000-4000$, this leads to an increase of $\Psi(h)$ with elongation of $h$ (see Fig.~3b--d).

It is worth emphasizing that the estimation formula (37) provides only an upper limit of the extrapolation uncertainties. Indeed, it involves scalar deviations $({\bf \tilde f}_i - {\bf f}_i)^2$ at the end of each outer interval $h$ without taking into account that the force ${\bf f}$ is a vector which can change its direction during inner time steps. Such a change may lead to a compensation of uncertainties and, thus, to their decrease. The fact that Eq.~(37) overestimates the extrapolation errors is confirmed in Fig.~3, where we see that $\Psi(h)$ does not fall to negligible values even at a tiny outer time step of $h = 12$~fs, while $\lim_{h \to 0} \Psi(h)=0$ by definition. Instead, all the dependencies $\Psi(h)$ in Fig.~4 tend to a some finite level of $\Psi_0 \sim 2-3\%$ when $h$ approaches very small values. Thus, the most simplest way to correct the estimation given by Eq.~(37) is to extract $\Psi_0$ from $\Psi$, i.e., $\Delta \Psi = \Psi - \Psi_0$. More accurate estimations could be to calculate the deviations at each inner time step. But this will require enormous computational costs which are significantly larger than those needed for the extrapolation of forces itself, making no sense to perform the estimations of such a kind.

\textit{Investigation of conformational properties.} --- Estimations of the extrapolation accuracy made in the preceding subsection will now be confirmed in actual investigations of conformational properties. To accomplish this we consider two systems, namely, hydrated alanine dipeptide and miniprotein in aqueous solution. In the first case we study the dipole moment distribution of the solute molecule. The second one is devoted to protein folding.

The dipole moment distribution functions $Q(\zeta)$ of the hydrated alanine dipeptide molecule obtained in MTS-MD/OIN/3D-RISM-KH simulations using the SFE, ASFE, GSFE$'$, GSFE, and ESFE extrapolation methods are shown in Fig.~4. For the purpose of comparison, ``exact'' data related to conventional MD (CMD) with explicit solvent are also included there. Note that function $Q(\zeta)$ presents the probability for the system to stay in a microscopic state with dipole moment $\zeta$. Because of this, it is normalized, $\int Q(\zeta) d \zeta = 1$, where $\zeta=|\bm{\zeta}|$ is the magnitude of the dipole moment $\bm{\zeta}=\sum_i q_i {\bf r}_i$ of the solute macromolecule satisfying the electro-neutrality condition $\sum_i q_i = 0$. Such a probability is very sensitive to the choice of solute and solvent models \cite{Kwac, Ishizuka} as well as to any uncertainties in the force evaluations. Therefore, a comparison of $Q(\zeta)$ with its ``exact" counterpart is a good idea for testing any new approach. The ``exact" (or rather ``expected") values of $Q(\zeta)$ were calculated with tiny time steps $\delta t=1$~fs and $\Delta t=4$~fs without any extrapolation ($h=\Delta t$) to minimize the influence of all possible numerical uncertainties on the results.

\begin{figure}[t]
\centerline{\includegraphics[width=0.65\textwidth]{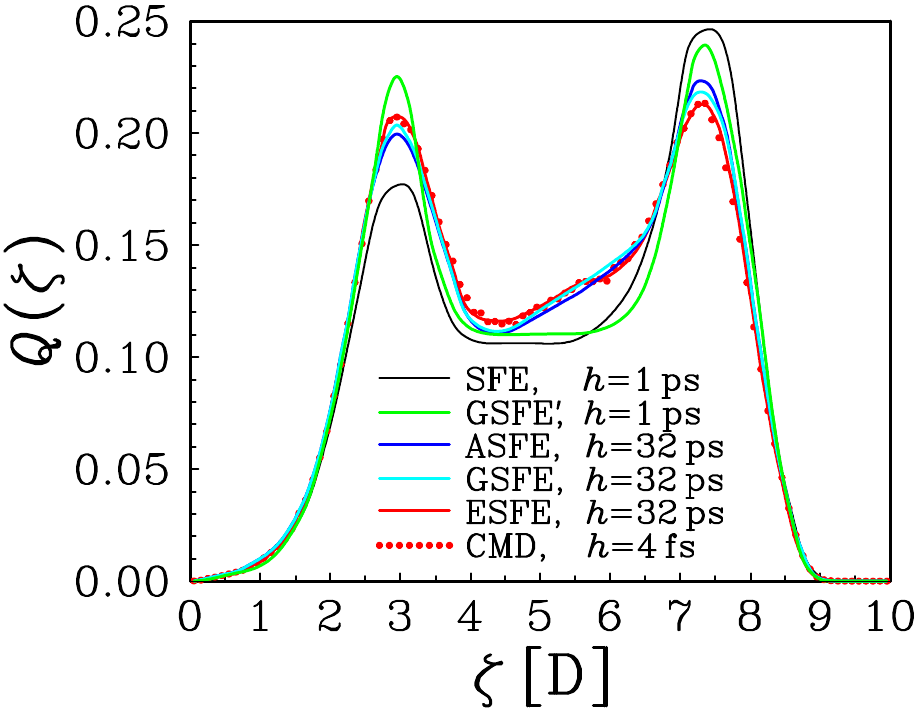}}
\caption{Dipole moment distribution of hydrated alanine dipeptide molecule obtained in MTS-MD/OIN/3D-RISM-KH simulations using different extrapolation approaches with different outer time steps in comparison with ``exact'' data (see the text).}
\end{figure}

In the ``exact" distribution function we can observe two clear peaks at $\zeta \approx 3$~D and $\approx 7.5$~D including some enhancement in intermediate region $\zeta \approx 5.5$~D. This corresponds to different conformational states of the alanine dipeptide molecule. A similar behavior of $Q(\zeta)$ was established earlier for various force fields and water solvent models, and was compared with experimental (infrared spectroscopy) results for the real system of hydrated alanine dipeptide. \cite{Luchko:2010:6:607, Kwac} As we can see, the SFE and GCFE$'$ schemes are not able to reproduce these features qualitatively even at a relatively small outer time step of $h=1$~ps. Here the deviations from the ``exact" data are significant, especially for SFE. The accuracy increases considerably when going to the ASFE and GSFE approaches even through a huge outer time step of $32$~ps is used. Nevertheless, the uncertainties are still visible here although they are small. Only the ESFE curve at $h=32$~ps is indistinguishable from the CMD data in the whole $\zeta$-range. Therefore, as was theoretically predicted by us above on the basis of the $\Psi(h)$-behaviour (Figs.~2 and 3), the deviations decrease when ordering the extrapolation methods in the following sequence: SFE, GSFE$'$ or ASFE, GSFE, and ESFE.

In the second example we consider the ability of the new approach to study protein folding. The corresponding MTS-MD/OIN/ESFE/3D-RISM-KH simulations were carried out at a size of the outer time step of $h=4$~ps using the same force field and parameters described at the top of this section. The only difference is that now the temperature of the system was increased from $T=300$ to 325~K to be consistent with previous investigations by the generalized Born \cite{Simmerling:2002:124:11258} and GSFE \cite{Omelyann} approaches. Moreover, the simulations started ($t=0$) from a well denatured configuration. We used the new cartoon representation with STRIDE \cite{Frishman:1995:23:566} in the VMD (Visual Molecular Dynamics) package \cite{Humphrey:1996:14:33} in which the secondary structure formations are assigned as follows: $\alpha$-helix (purple), $\beta$-sheet (yellow), turn (cyan), coil (white), and $3_{10}$-helix (blue).

\begin{figure}
\centerline{
\includegraphics[width=0.68\textwidth]{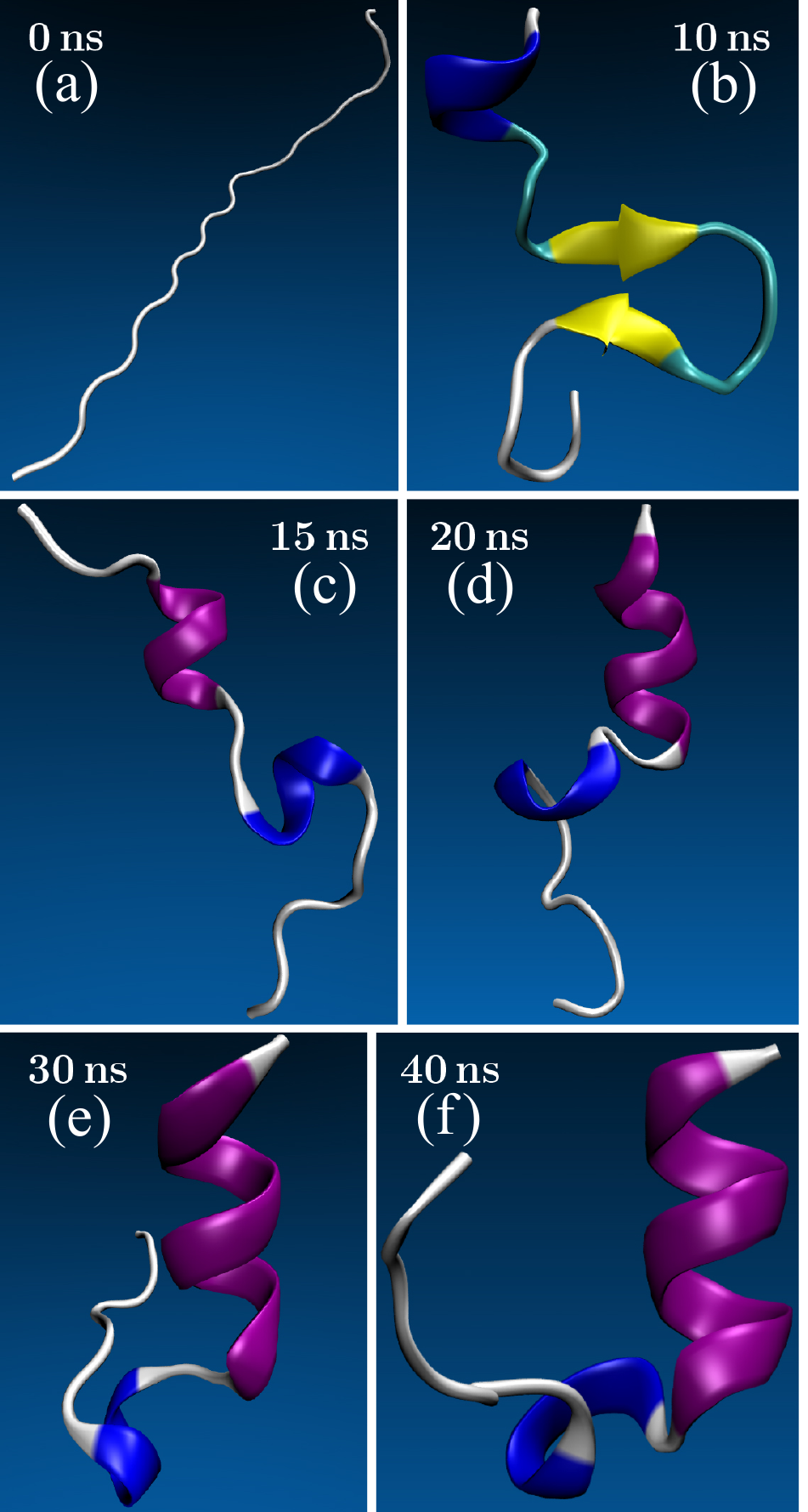}}
\caption{Conformational states of hydrated miniprotein 1L2Y in tertiary structure representation, obtained during the MTS-MD/OIN/ESFE/3D-RISM-KH simulatiuons at different moments of time: (a) $t=0$, (b) $t=10$~ns, (c) $t=15$~ns, (d) $t=20$~ns, (e) $t=30$~ns, and (f) $t=40$~ns.}
\end{figure}

\begin{figure}[ht]
\centerline{
\includegraphics[width=0.72\textwidth]{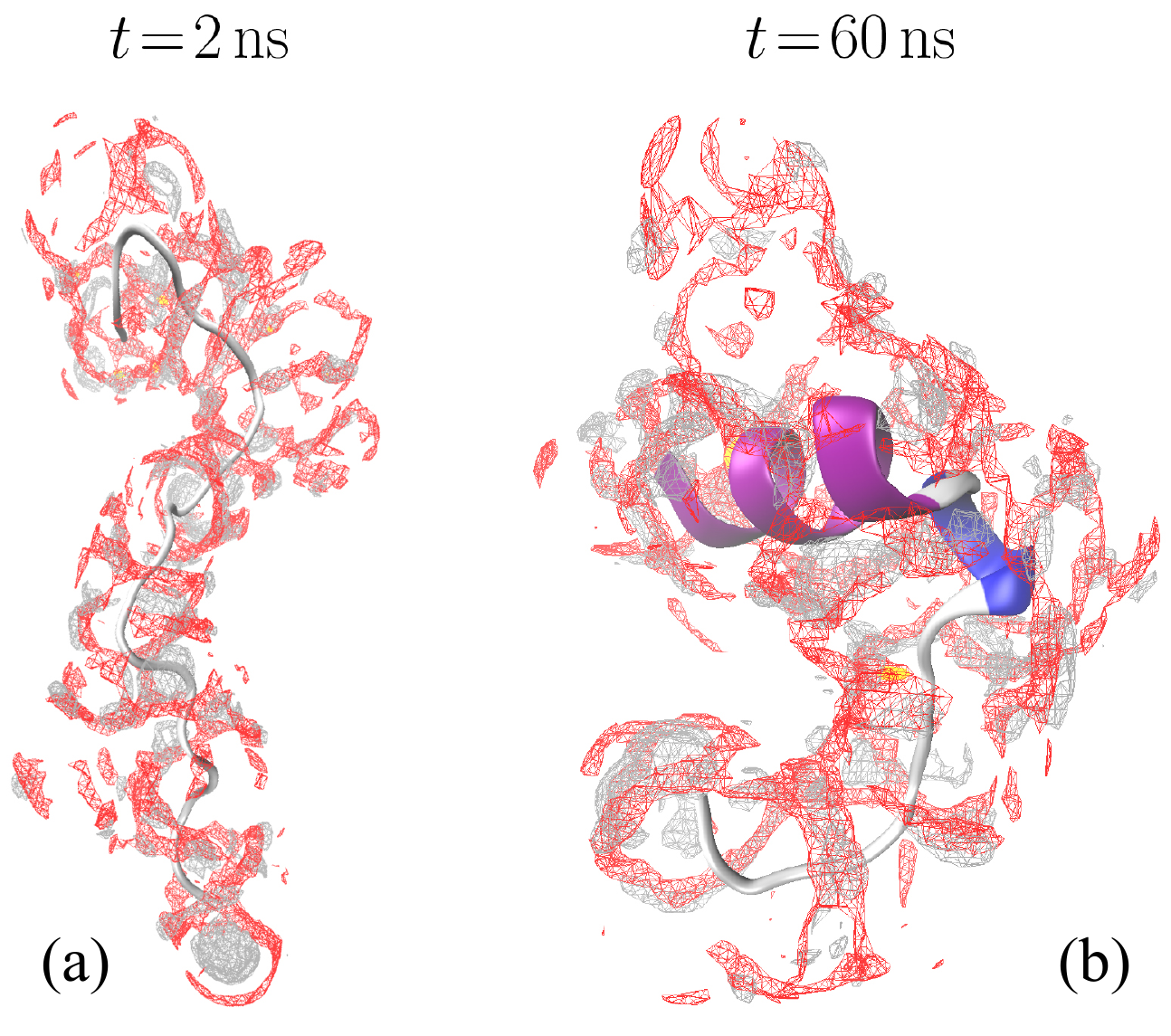}}
\caption{Tertiary structure snapshots of miniprotein 1L2Y at the (a) beginning ($t=2$~ns) and (b) end ($t=60$~ns) of the OIN/ESFE/3D-RISM-KH quasidynamics complemented by the isosurfaces of density distribution functions of water oxygen at $g_{\rm O}^{\rm uv}=3$ (red) and hydrogen at $g_{\rm H}^{\rm uv}=2$ (silver) together with spots of water oxygen with $g_{\rm O}^{\rm uv} > 8$ (yellow).}
\end{figure}

Six types of the tertiary structure of hydrated miniprotein obtained in our MD/OIN/ESFE/3D-RISM/KH approach are presented in Fig.~5. They correspond to different simulation lengths, namely, $t=0$, $10$, $15$, $20$, $30$ and 40~ns. As can be seen, at $t=10$~ns the miniprotein exhibits a misfolding (aggregated) behaviour with the presence of $\beta$-sheets and incorrectly placed (with respect to the native state) $3_{10}$-helix formation. Soon at $t=15$~ns, the $\alpha$-helix and correct $3_{10}$-helix structures arise instead. With the course of time at $t=15$ and $20$~ns they are extended in full, staying more and more close to their original forms. Already at $t=30$ and $40$~ns we can say about almost folded conformations which only slightly differ from that of the completely folded state. The latter is achieved nearly at $t=60$~ns, where the atomic root-mean-square deviations do not exceed about $1$~\AA\ with respect to the native configuration taken from the PDB of 1L2Y. \cite{Neidigh:2002:9:425} The tertiary structure of hydrated miniprotein obtained at the end of the simulations at $t=60$~ns is shown in Fig.~6 together with isosurfaces of the density distribution functions $g_\alpha^{\rm uv}({\bf r})$ of water oxygen ($\alpha={\rm O}$) and hydrogen ($\alpha={\rm H}$). The isosurfaces related to a denatured configuration at the beginning ($t=2$~ns) are also plotted there for comparison. It should be mentioned that a similar folding behaviour of the hydrated miniprotein was observed earlier within MD/OIN/3D-RISM/KH using the GSFE scheme. However, a moderate size of the outer time step of $h=1$~ps was allowed to use there. Now, we was able to apply a much longer step of $h=4$~ps, significantly accelerating the simulations (see the next subsection).

\textit{Acceleration of simulations.} --- Speedup of the MTS-MD/OIN/3D-RISM/KH simulations was investigated in the case of hydrated miniprotein using the GSFE and ESFE approaches within the \verb"SANDER" module of the Amber package. \cite{Amber} Note this module has a limitation on the number of CPU cores which can be involved in parallel calculations. Namely, it cannot exceed the number of protein residues and should be a power of 2. Taking into account that the 1L2Y miniprotein with $M=304$ atoms constitutes a 20-residue amino acid sequence (within the so-called tryptophan cage TC5b), the original \verb"SANDER" module can be applied with no more than 16 cores. Because of this we have modified the code when implementing our new approach to have the possibility to involve a much larger number of processors. Now it restricted only to the number $M$ of atoms in the macromolecule, rather than to the number of residuals. All the calculations were performed on parallel clusters of WestGrid -- Compute Canada national advanced computing platform.

\begin{figure}[t]
\centerline{\includegraphics[width=0.72\textwidth]{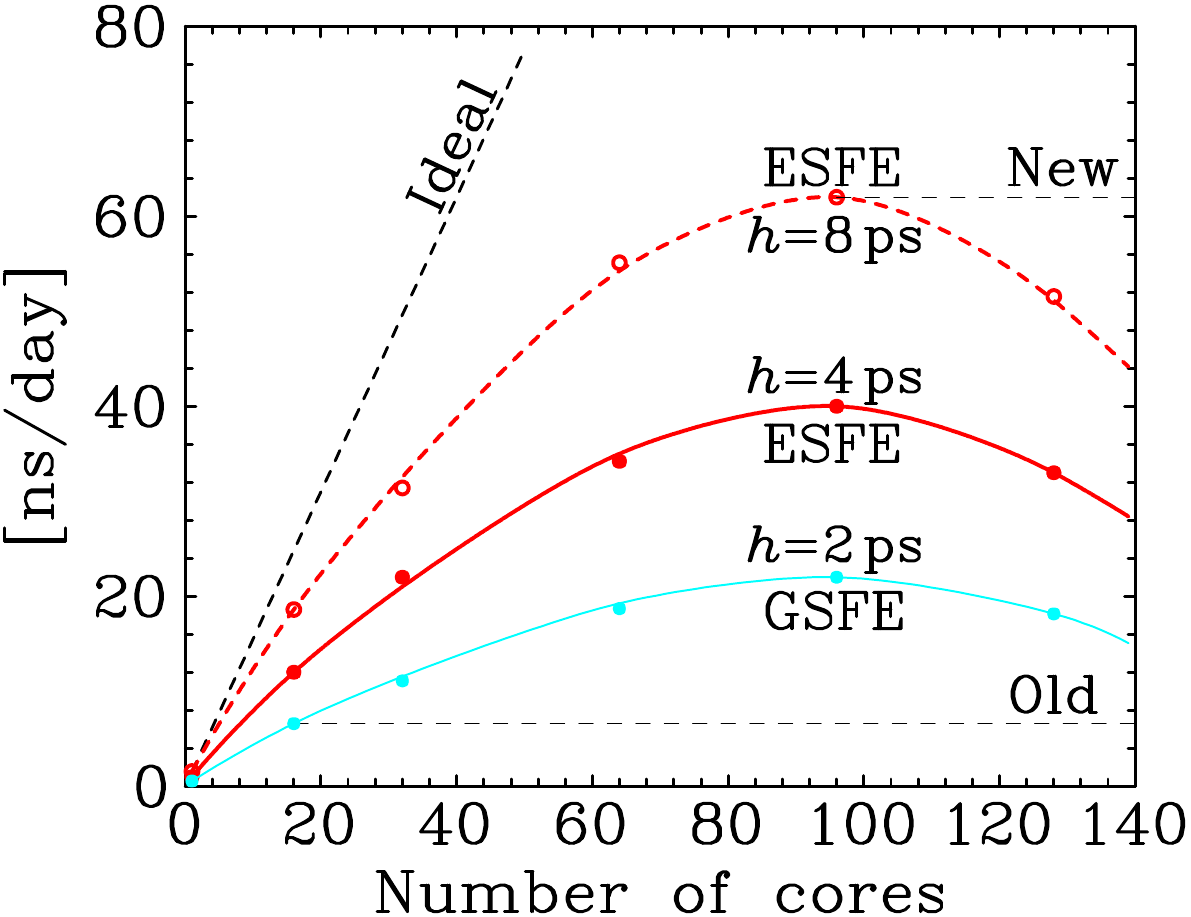}}
\caption{Productivity of the MTS-MD/OIN/3D-RISM-KH simulations within the GSFE ($h=2$~ps) and ESFE ($h=4$~ps and $h=8$~ps) approaches against the number of CPU cores.}
\end{figure}

Figure 7 shows the productivity $\mathcal{P}$ achieved in our simulations versus the number $\mathcal{N}$ of parallel CPU cores utilized at three fixed sizes of the outer time steps, namely, $h=2$, 4, and 8~ps. The first and third sizes correspond to the maximal steps allowed by the GSFE and ESFE approaches, respectively. Mention that at $h > 2$~ps, the GSFE uncertainties become too large, $\Psi \gtrsim 11\%$, while even a somewhat lower level of $\Psi \sim 10\%$ relates to the ESFE method at $h = 8$~ps, see Fig.~3c. The intermediate value $h=4$~ps should be considered as an optimal alternative for ESFE, where the precision ($\Psi \sim 7.7$\%) and productivity are both sufficiently high. Note also that we slightly decreased the solute-solvent truncation radius to $R_{\rm c}=10$\AA\ to reach an optimal performance with nearly the same precision. For the same reason, the reuse frequency was increased from $p=25$ at $h=2$ and $4$~ps to $p=50$ at $h=8$~ps. Six runs with $\mathcal{N}=1$, $16$, $32$, $64$, $96$, $128$ were carried out at each $h$. The corresponding values of $\mathcal{P}$ in these six points (shown as circles) were then taken as a basis to build smooth functions $\mathcal{P}(\mathcal{N})$ between and outside them with the help of a least-square spline procedure.

From Fig.~7 we see that using the previous GSFE approach and the old code with 16 cores lead to a productivity of 6.6 ns/day (lower lying horizontal dashed line). The improved code with 96 cores increases the efficiency more than in three times to a value of 22 ns/day. Further increase nearly in three times can be reached by applying the new extrapolation method ESFE which provides a productivity of 62 ns/day with 96 cores at $h=8$~ps (upper lying horizontal dashed line). Overall, this leads to the acceleration of the simulations in a factor of 10, enabling to quickly fold the miniprotein from a fully extended state spending only one day of the calculations (according to Figs.~5 and 6). Even at intermediate $h=4$~ps, the efficiency increases from 6.6 to 40 ns/day, i.e. in six times with respect to the previous GSFE scheme within the old core. All the three curves $\mathcal{P}(\mathcal{N})$ in Fig.~7 exhibit a saturation regime at $\mathcal{N} = 96$. At $\mathcal{N} > 96$ the performance decreases with increasing $\mathcal{N}$ due to time loss on interprocessor communications. Without this loss (on an ideal supercomputer in future) we could come to an ideal productivity (see the dashed line) which is a linear function $\mathcal{N}$ for any number of cores.

\section{Conclusion}

In this paper we have developed an enhanced approach to the extrapolation of solvation forces for speeding up hybrid MD/3D-RISM-KH simulations of complex biochemical systems. It extends and improves our previous approximated schemes by additionally incorporating new techniques into the extrapolation strategy. They include an exponential scaling transformation of coordinate space accompanied with an automatically adjusted balancing between the least square minimization of force deviations and the norm of coefficients in the approximation. The exponential scaling linearizes and smoothes the 3D-RISM-KH solvation forces, leading to an extra accuracy of the extrapolation. The dynamical balancing provides exact results in limits when the current spatial configuration is close to those belonging to the reference list. This is in a contrast to the earlier approaches which produce approximate values at any point of the configurational space. Other techniques, such as individual non-Eckart transformations (to properly account changes of the solvation forces caused by local rotations of segments of the solute macromolecule) and an extension of the reference list (to choose the best subset of basic configurations), are also involved into the new approach.

The expensive 3D-RISM-KH solvation forces were expressed in terms of microscopic interaction potentials between solute and solvent atoms via the quasiequilibrium density distribution functions of solvent atoms around the solute biomolecule in its current conformation. During the dynamics these forces are explicitly calculated only after every long enough (outer) time interval, i.e., quite rarely to reduce the computational costs. At much shorter (inner) time steps, these forces are extrapolated on the basis of their outer values taken from the reference list. The equations of motion are then solved using a multiple time step integration (MTS) within an optimized isokinetic Nos\'e-Hoover (OIN) chain thermostat. The new enhanced method has been applied to MTS-MD/OIN/3D-RISM-KH simulations of different solvated organic and biomolecular systems including proteins. It has been demonstrated that the enhanced extrapolation allows one to achieve much better accuracy in the solvation force approximation than the existing approaches. As a result, it can be used with much larger outer time steps, leading to a significant acceleration of the simulations. For instance, a speedup in several times up to one order of magnitude is possible in the case of miniproteins.

The new approach can be applied to more complicated proteins and other biomolecular and biochemical systems, biomaterials, including cellulose nanocrystals, in different solvents and electrolyte solutions. It can be also combined with the replica exchange method and implemented into the current version of Amber. These and other topics will be the subject of our future studies.

\section*{Acknowledgments}

The computations were carried out on the high performance computing resources provided by WestGrid of Compute/Calcul Canada.

\end{document}